\documentclass[a4paper,11pt]{article}
\pdfoutput=1
\usepackage{graphicx}
\usepackage{xcolor}

\newcommand{\Fc}{\mathcal{F}}
\newcommand{\Gc}{\mathcal{G}}
\newcommand{\Kc}{\mathcal{K}}

\usepackage[english]{babel}
\usepackage{amssymb,graphicx}
\usepackage{amsmath,amsfonts}
\usepackage{fullpage}

\title{Non-local self-healing of Higgs inflation}
\author{Alexey S. Koshelev$^{1}$ and Anna Tokareva$^{2}$
\\ \mbox{}$^{1}$\textit{\small Departamento de F\'isica, Centro de Matem\'atica e Aplica\c{c}\~oes (CMA-UBI),}
\\\textit{\small Universidade da Beira Interior, 6200 Covilh\~a, Portugal}
\\ \mbox{}$^{2}$\textit{\small Institute for Nuclear Research of Russian Academy of Sciences, 117312 Moscow, Russia}
}

\date{~}
\begin{document}
\maketitle

\begin{abstract}
Higgs inflation is known to be a minimal extension of the Standard Model allowing for the description of the early Universe inflation. This model is considered as an effective field theory since it has a relatively low cutoff scale, thus requiring further extensions to be a valid description of the reheating phase. We present a novel unified approach to the problem of unitarization and UV completion of the Higgs inflation model without introducing new massive degrees of freedom. This approach is based on an analytic infinite derivative modification of the Higgs field kinetic term. We construct a unitary non-local UV completion of the original Higgs inflation model while the inflationary stage is kept stable with respect to quantum corrections.

\end{abstract}

\section{Introduction}

The minimal realization of the inflationary stage in the early Universe can be achieved by making use of a scalar field which is already present in the Standard Model -- the Higgs field \cite{Bezrukov:2007ep}. If the Higgs boson has a non-minimal coupling to gravity as follows\footnote{We adopt the $(-+++)$ signature convention for the metric in the paper.}
\begin{equation}
\label{action}
S=\int d^4 x\left(\frac{1}{2}(M_P^2+2 \xi H^{\dagger}H )R- |D_{\mu}H|^2 - \frac{\lambda}{4} (H^{\dagger}H-v^2)^2\right)\,,
\end{equation}
it can successfully drive inflation. Notations are standard and quite self-explanatory but several comments are in order here. In this model, the predictions for the cosmological perturbations are perfectly consistent with the recent data \cite{Akrami:2018odb}. However, the Higgs self-coupling is fixed by the low energy physics in such a way that at high energies it cannot be too small. If no new physics enters at energies between the Planck and electroweak scales the value of $\lambda$ at the scale of inflation should be of order $10^{-2}$. This requires a large value for the dimensionless coupling $\xi\sim 10^3\div 10^4$ in order to obtain the observed amplitude of cosmological perturbations. This leads to the problems of the theoretical consistency of this inflationary model due to the low value of the cutoff scale $M_P/\xi$ \cite{Burgess:2009ea,Barbon:2009ya} which is higher than the scale of inflation but is quite close to it. Although inflation itself can still be consistently described \cite{Bezrukov:2010jz}, during the reheating phase the inflaton produces particles with momenta higher than the cutoff scale \cite{Ema:2016dny}. This means that the model enters a strongly coupled regime and a UV completion is required. {In principle, the model could contain a kind of self-healing mechanism and can still work in a non-perturbative regime {at high energies} without violating unitarity \cite{Calmet:2013hia,Ema:2020zvg}. However, the renormalizability problem remains in this model and as the consequence the predictivity of such a model is lost, {in a sense of obtaining concrete predictions from the values of parameters measured at low energies}.}

Several models which complete Higgs inflation at energies higher than $M_P/\xi$ were suggested \cite{Giudice:2010ka,Ema:2017rqn,Barbon:2015fla,Gorbunov:2018llf,He:2018gyf,He:2018mgb}. All these models assume the presence of an extra scalar particle with the mass of the order of $M_P/\xi$. Moreover, in all these models, inflation is actually driven by this extra field instead of the Higgs boson. However, the background energy density is dominated by the Higgs field and the amplitude of scalar perturbations is still defined by $\lambda$ and $\xi$. Therefore it is interesting to address a question of whether we can build a UV completion for the Higgs inflation mechanism without introducing new degrees of freedom.

{Presence of gravity in this setup renders the Higgs inflation model non-renormalizable in its original formulation.} This means that the model contains an infinite number of operators suppressed by the powers of the cutoff scale. The coefficients standing in front of these operators are to be defined from the experiment. Notice that besides corrections to the inflaton potential, in general, we expect also operators with (infinitely) higher derivatives. The latter observation gives us a hint that higher derivative or even non-local effects can be relevant in this construction. One can ask whether it is possible to tune all coefficients of the effective Lagrangian in such a way that the model would be still predictive at energies higher than the cutoff scale? In the current paper, we show that the answer is positive and formulate an Analytic Infinite Derivative (AID) Lagrangian leading to the finite loop corrections which also might in principle solve the strong coupling issue in Higgs inflation.

AID Lagrangians were widely studied in various contexts ranging from string field theory \cite{Witten:1985cc,Witten:1986qs,Ohmori:2001am,Arefeva:2001ps} and in particular p-adic string theory \cite{Brekke:1988dg,Vladimirov:1994wi,Dragovich:2017kge} to the models aimed at describing quantum gravity \cite{Kuzmin:1989sp,Krasnikov:1987yj,Tomboulis:1997gg,Biswas:2011ar,Biswas:2016egy,Koshelev:2017ebj}. An interacting scalar field theory can be rendered finite upon introducing an appropriate modification to the quadratic in fields term. Presenting sketchy,
the Lagrangian of the form
\begin{equation}
\label{lagrangian}
L= \frac{1}{2}\phi \Fc(\square)\phi - V(\phi)\,,
\end{equation}
with $\Fc(\Box)=(\square-m^2)\, e^{2\sigma(\square)}$ can lead to the exponential suppression of the propagator at large momenta as long as a suitable choice of function $\sigma(\square)$ in the exponent is made. This yields all loop integrals to become finite.

For the time being the most robust motivation for theories with an infinite number of higher derivatives in the form of analytic form-factors comes from the string field theory. Namely, the kind of Lagrangians which arise in such an approach looks as follows \cite{Ohmori:2001am,Arefeva:2001ps}:
\begin{equation}
\label{lSFT}
L_{\rm SFT}= \frac{1}{2}\varphi (\square-m^2)\varphi - V(e^{-\sigma(\square)}\varphi)\,,
\end{equation}
where $\sigma(\square)$ is some polynomial of the d'Alembertian $\square$ and $V$ is some interaction, polynomial in its argument and the lowest degree of the field $\varphi$ in $V$ is 3. In other words, $V$ does not contain quadratic in $\varphi$ terms. Technically $V$ is not an interaction potential anymore since it has clear momentum dependence. However, given $\sigma(\square)$ is a polynomial we have a well-defined IR limit and the real potential is extracted when all the momenta are zero. To discriminate the regimes of low and high energies it is natural to setup a mass scale $\Lambda$ of the higher derivative modification and write $\sigma(\square)$ such that it depends on $\square/\Lambda^2$. In string theory, this scale is the string mass which is theoretically bounded by the Planck mass $M_P$ from above.

From the perspective of the latter Lagrangian we have a theory with AID vertices. One can however easily redefine the field as $\phi=e^{-\sigma(\square)}\varphi$ and move the AID operators to the quadratic in fields term.
This yields
\begin{equation}
\label{lSFT2}
L= \frac{1}{2}\phi (\square-m^2)e^{2\sigma(\square)}\phi - V(\phi)\,,
\end{equation}
which is exactly Lagrangian (\ref{lagrangian}).

In a well-defined scenario, one should avoid ghost fields and this is one of the guiding principles limiting our AID theory construction.
We know starting from the papers by Ostrogradski \cite{Ostro:1850} that higher derivatives generically introduce ghosts. Since the number of degrees of freedom is counted by the number of poles in the propagator one may try to keep only one pole even with extra derivatives in the Lagrangian. Given the construction above this can be made only if the original operator $(\square-m^2)$ in the quadratic form is multiplied by a function of the d'Alebertian which has no zeros on the whole complex plane.
Mathematically the only possibility for such an extra factor is an exponent of an entire function. This being said as long as the ghost absence question is concerned we can consider $\sigma(\square)$ to be a generic entire function and not only a polynomial.
The presented construction can be trivially elevated to models in a curved background by a simple replacement of the flat space-time d'Alembertian by its covariant counterpart. This will not spoil the ghost-free condition in any way.

We stress however that the non-local function itself depends on the particular vacuum of the potential in order to guarantee the absence of ghosts. Namely, consider a potential that has several vacua which moreover arrange different masses to the scalar field. The presented above construction allows having no ghosts only in a single vacuum in which the mass of the field is given by $m^2$. In all vacua where the effective mass of the field is different from $m^2$ an infinite number of fields with complex masses appear on top of the standard excitation.
An interpretation of these new fields is still unclear, while there are several including reasonably old studies \cite{Coleman:1968wh,Hawking:2000bb} claiming that they do not spoil unitarity at least at scales below the scale of higher-derivative modification $\Lambda$. Another approach designates such fields as totally virtual degrees of freedom and a term fakeon is used with respect to such excitations \cite{Anselmi:2018kgz} especially in describing certain aspects of Lee-Wick type models.

The simplest choice for an entire function is a polynomial, and the minimal try would be $\sigma(\Box)=-\Box/\Lambda^2$. This minimal choice, however, appears to lead to the exponential growth of tree-level scattering amplitudes given by the exchange diagrams which can be shown by direct simple computation. On the other hand, the choice $\sigma(\Box)=\Box^2/\Lambda^4$ does not result in such a problematic behavior providing us with a UV finite theory even if the interaction potential is non-renormalizable in the original local model. {The difference comes from the fact that in the latter case the exponent is decaying along both real and complex axis and necessity of this was noted already in \cite{Tomboulis:1997gg}.} Such kind of theories can become strongly coupled at energies higher than the non-locality scale. But due to UV finiteness, the predictivity would still hold because only a finite number of parameters is present unlike the case of other self-healing mechanisms \cite{Calmet:2013hia}. In order to extract predictions, one may need non-perturbative methods, resummation of diagrams, lattice computations, etc. Nevertheless, the good point here that this is possible in principle and this is why such analytic non-local models also considered as a possible approach to quantize gravity.

In this paper, we exploit this kind of analytic non-locality in order to heal the Higgs inflation model from the problem of {non-renormalizability. We show that the non-local propagator can be chosen in such a way that all loop diagrams become convergent. We show that, at least at tree level, the model can be made weakly coupled even for momenta larger than the scale of non-locality (the suppression scale of higher derivatives). Nevertheless, we leave the rigorous loop computations for future study, restricting ourselves with the computation of the one-loop effective potential for the real scalar field with non-local propagator. We show that if the non-locality scale is chosen to be of the order of the cutoff scale of the initial local scalar field Lagrangian, the one-loop correction given by the Coleman-Weinberg potential can be small for all values of the classical background field. This provides a hint that the model can be still weakly coupled also at the loop level. }

The paper is organized as follows. In Section~2 we setup a non-minimally coupled scalar field model of inflation with an AID propagator. We proceed with outlining the main steps in constructing so to say good AID theories and give a special account to the problem of extra modes effectively appearing as long as the system leaves the vacuum. We further show that inflation remains stable with respect to quantum corrections. Having this done we move on in Section~3 by incorporating the model in the Standard Model Higgs inflation scenario paying special attention to the issues of preserving the gauge invariance, unitarity of amplitudes, and the problem of the Higgs mass naturalness. The results are summarized in Section~4.

%%%%%%%%%%%%%%%%%%%%%%%%%%%%%%%%%%%%%%%%%%%%%%%%%%%%%%%%%%%%%%%%%%%%%%%%%%%%%%%%%%%%%%%%%%%%
\section{UV completion of a toy model: real scalar field with non-minimal coupling to gravity.}

In this Section, we demonstrate distinctive features of our approach using a simple example of a single scalar field non-minimally coupled to gravity. This field would resemble the behavior of the radial Higgs mode. A way of introducing non-locality is presented and critically considered from several points of view.
%analyzed from and its implications for the rest of the Standard model will be discussed in the next Section.

\subsection{The toy model}

We start with the following action
\begin{equation}
S=\int d^4 x\left(\frac{1}{2}(M_P^2+\xi h^2 )R- \frac{1}{2}(\partial_{\mu}h)^2 - \frac{\lambda}{4}h^4\right)\,.
\end{equation}
After the redefinition of the metric we obtain an action in the Einstein frame \cite{Bezrukov:2007ep}
\begin{equation}
\label{EFlagrangian}
S_E=\int d^4x\sqrt{-g_E}\left(\frac{1}{2}M_P^2 R_E- \frac{1}{2}(\partial_{\mu}\phi)^2 - V(\phi)\right)\,,
\end{equation}
where
\begin{equation}
\label{potential}
V(\phi)=\frac{\lambda M_P^4 \,h(\phi)^4}{4(M_P^2+\xi h(\phi)^2)^2}~\text{ and }~\frac{d \phi}{dh}=\frac{M_P\sqrt{M_P^2+(1+6\xi)\xi h^2}}{M_P^2+\xi h^2}\,.
\end{equation}
For $h>M_P$ the potential simplifies to
\begin{equation}
V(\phi)\approx\frac{\lambda M_P^4}{4\xi^2}(1-e^{-2\phi/(\sqrt{6}M_P)})^2\,,
\end{equation}
which resembles the one of the Starobinsky model of inflation \cite{Starobinsky:1980te,Starobinsky:1981vz,Starobinsky:1983zz}.

We continue working in the Einstein frame Lagrangian for the Higgs inflation model and also at this point we introduce the non-local propagator yielding the following Lagrangian:
\begin{equation}
\label{NLlagrangian}
L=\frac{1}{2}M_P^2 R_E+ \frac{1}{2}\phi \Fc(\square)\phi - V(\phi)\,.
\end{equation}
As explained in the Introduction, we choose the operator function
\begin{equation}
\Fc(\Box)=\Box\, e^{\square^2/\Lambda^4}\,,
\label{0F}
\end{equation}
as the simplest possibility which arranges at once the absence of new degrees of freedom, suppression of all loop integrals, and does not lead to the growth of tree amplitudes. For our purposes of constructing UV completion at the cutoff scale $M_P/\xi$ we need the non-locality scale to obey $\Lambda\sim M_P/\xi$. We leave the potential unmodified compared to (\ref{potential}), in order to get the original model at the scales below the scale of $\Lambda$.

We follow the canonical understanding that the number of degrees of freedom is given by the number of finite poles of the propagator. Our choice for $\Fc(\square)$ does satisfy the ghost-free condition as long as $V''=0$. The latter condition is indeed satisfied in both vacua of the potential in question (\ref{potential}), namely, when $\phi$ tends to infinity and when $\phi=0$. In some sense, it is a lucky situation that the desired potential generates the same masses of a particle in all its vacua. In our case the particle is massless. Otherwise, given mass terms are different in different vacua, the form-factor would be able to arrange an absence of ghosts only in one of the vacua. It follows from here that it is a good thing that the field potential of our model has no other (otherwise irrelevant) vacua. The existence of such other vacua and most probably other non-zero value of mass terms for the scalar field in them could lead to some non-perturbative effects related to possible violations of the ghost-free condition. We comment on this issue in more detail right below.

%%%%%%%%%%%%%%%%%%%%%%%%%%%%%%%%%%%%%%%%%%%%%%%%%%%%%%%%%%%%%%%%%%%%%%%%%%%%%%%%%%%%%%%%%%%%
\subsection{On the number of physical excitations}

As noted above, all two existing scalar field vacua in our model obey $V''=0$ and therefore operator function (\ref{0F}) generates only one degree of freedom.
As long as $V''\neq 0$ which is true almost everywhere apart from the vacua an attempt to linearize equations results in the following operator in the quadratic form
\begin{equation}
\Kc_m(\square)=\frac12 \square e^{\square^2/\Lambda^4}-\frac12 m^2\,.
\label{manyroots}
\end{equation}
Here we use $m^2=V''(\phi)$ with $\phi$ being the classical background value of the scalar field. This particular operator results in infinitely many effective local scalar fields with most of them having complex masses.
We name these new fields effective because they do not belong to any vacuum of the model in a sense that they cannot be created as states in any present vacuum of the theory. Still, it is nothing wrong with linearizing around a non-vacuum point of evolution especially if the effective mass is a slowly varying quantity as it is during inflation.

To understand why many fields appear it is useful to represent the operator function of the d'Alembertian utilising the Weierstrass product decomposition for an entire function which tells us that any entire function $\Gc(z)$ can be presented as:
\begin{equation}
\Gc(z)=\prod_i(z-z_i)^{n_i}e^{g(z)}\,,
\label{wp}
\end{equation}
where $z_i$ are algebraic roots of the equation $\Gc(z)=0$, $n_i$ is the root's multiplicity and $g(z)$ is some entire function. In a simplest case all $n_i=1$. The outlined above operator $\Kc$ is a manifestly entire function. {A further justification to make use of this formula in general comes from the fact that at least the string field theory obtained models certainly contain only entire functions of $\square$ as operators in field quadratic forms.}
Therefore, in a generic case for models containing the following quadratic in field term
\begin{equation}
L=\frac12\phi\Gc(\square)\phi+\dots\,,
\label{lwp}
\end{equation}
with an entire operator function $\Gc(\square)$ one can easily achieve two things \cite{Koshelev:2007fi,Koshelev:2010bf}:
\begin{itemize}
\item
First, the free equation of motion can be easily solved. Indeed, it would look like
\begin{equation}
\Gc(\square)\phi=\prod_i(\square-m_i^2)e^{g(\square)}\phi=0\,.
\label{lwpeom}
\end{equation}
We assume here for simplicity that the root's multiplicity is always one. Then the solution will be
\begin{equation}
\phi=\sum_i\phi_i~\text{ where }~(\square-m_i^2)\phi_i=0\,.
\label{lwpeomsol}
\end{equation}
Moreover, since the original model provides a form-factor with real coefficients in its Taylor expansion around zero, all roots are either real or come in complex conjugate pairs.

For a pair of complex conjugate roots, numbered say $i$ and $j$, one should consider correlated initial conditions on functions $\phi_i$ and $\phi_j$ such that the sum of these functions is zero. This is a condition for a consistent background solution for the original field $\phi$ which must be real.

We note here that more than one real $m_i^2$ will definitely be a ghost-like excitation and is as such a problematic configuration. However, pairs of complex conjugate masses squared are not obligatory bad and may lead to hassle-free models. Notice that in this case one encounters somewhat non-canonical complex field model which has the following Lagrangian:
\begin{equation}
L_i=\frac12\left[\Gc'(m_i^2)\phi(\square-m_i^2)\phi+\Gc'({m_i^2}^\ast)\phi^\ast(\square-{m_i^2}^\ast)\phi^\ast\right]\,,
\label{lphiphistar}
\end{equation}
which cannot be diagonalized in real fields.
Factors of $\Gc'(m_i^2)$ appear upon computing the Weierstrass decomposition.

The appearance of complex conjugate poles and their interpretation was already discussed in \cite{Coleman:1968wh}. In a nut-shell, such poles with masses $m=\,u\pm i\nu$ with $\nu>0$ would imply causality violation at distances less than $1/\sqrt{\nu}$. Even generically an alarming symptom, it can be safely ignored given that the imaginary part which would cause troubles is large enough compared to physically important scales of the model. On top of this absence of classical growing modes should be guaranteed to claim safely that newly appeared particles do not interfere with the rest of the model. It is important to verify that this wishful expectation holds.
\item
Second, one can straightforwardly compute residues at all poles of the propagator. The point is to figure out by use of formula (\ref{wp}) that a residue at the point of $m_i^2$ is given by $1/\Gc'(m_i^2)$ which is not zero by construction as long as all roots are assumed to be the simple ones.

One technically important point here is that in principle fields can be rescaled to a somewhat arbitrary number. This implies that unless we clearly understand the field's normalization, or unless there are no eternal guiding principles for doing that, one can always bring the real part of the residue value to be $\pm1$.
\end{itemize}

The novel and an intuition breaking thing here is that the very situation of changing the number of degrees of freedom dynamically is extremely unusual and is met here only due to higher derivatives. Moreover, the jump is bizarre from a single excitation to effectively infinitely many of them and most of those extra ``would be'' excitations have complex masses which are totally alien objects in canonical field theory. What is even more curious, neither vacuum of the presented theory can create anything but one real massless particle.

Hence, in principle, we can say that those effective fields are not excitations in any way and as such just stop discussing them. However, since their appearance, even effective, is not a completely understood process, we are going to follow a safer way and show that these ``would be'' degrees of freedom are screened by having huge masses compared to real physical scales in the model. This follows from the adjustment of the non-locality scale $\Lambda$ to be heavier than the {Hubble scale during inflation.} Also we will formulate a condition allowing no classical growing modes for these new modes.

\begin{figure}[h!]
\center{
\includegraphics[width=7cm]{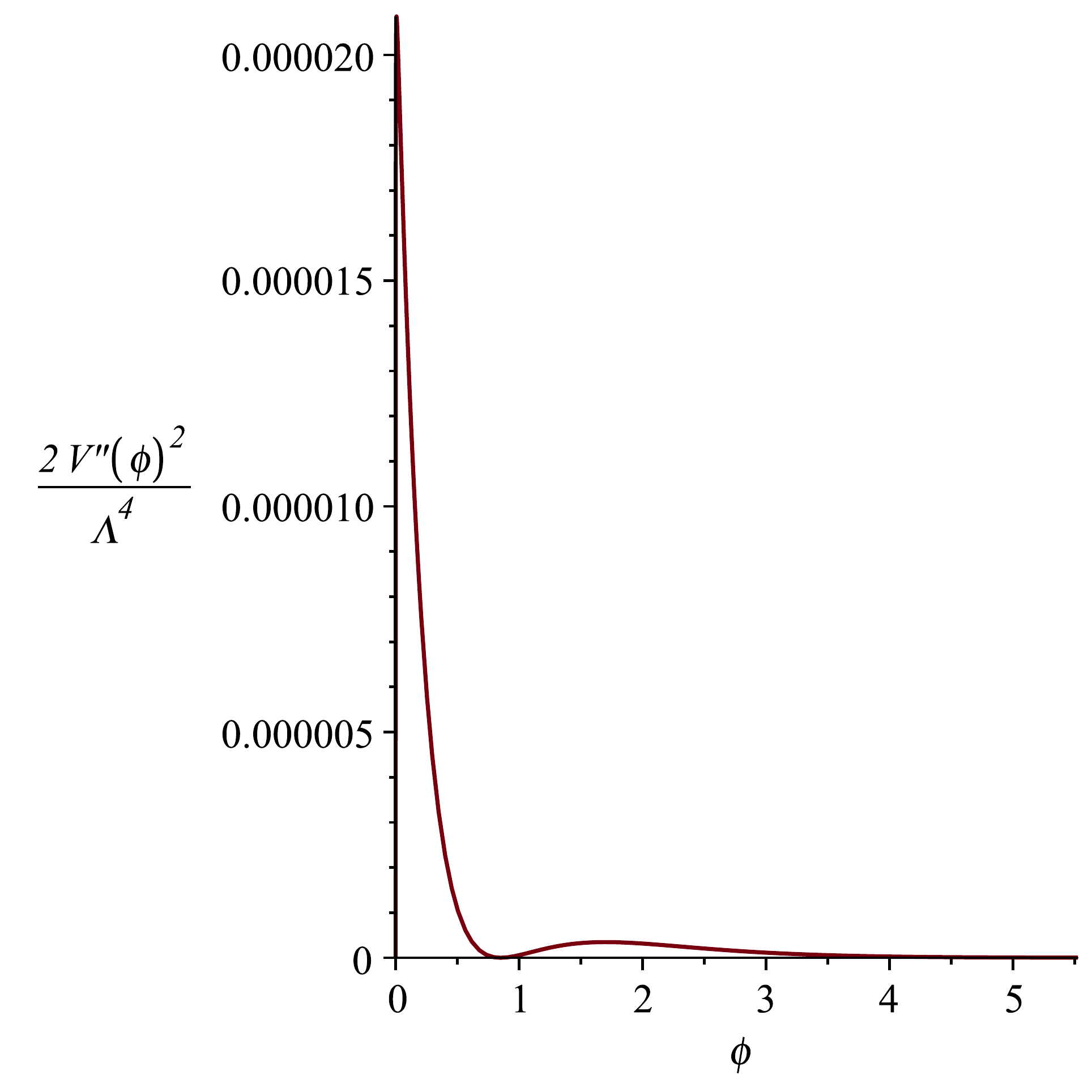}
\includegraphics[width=7cm]{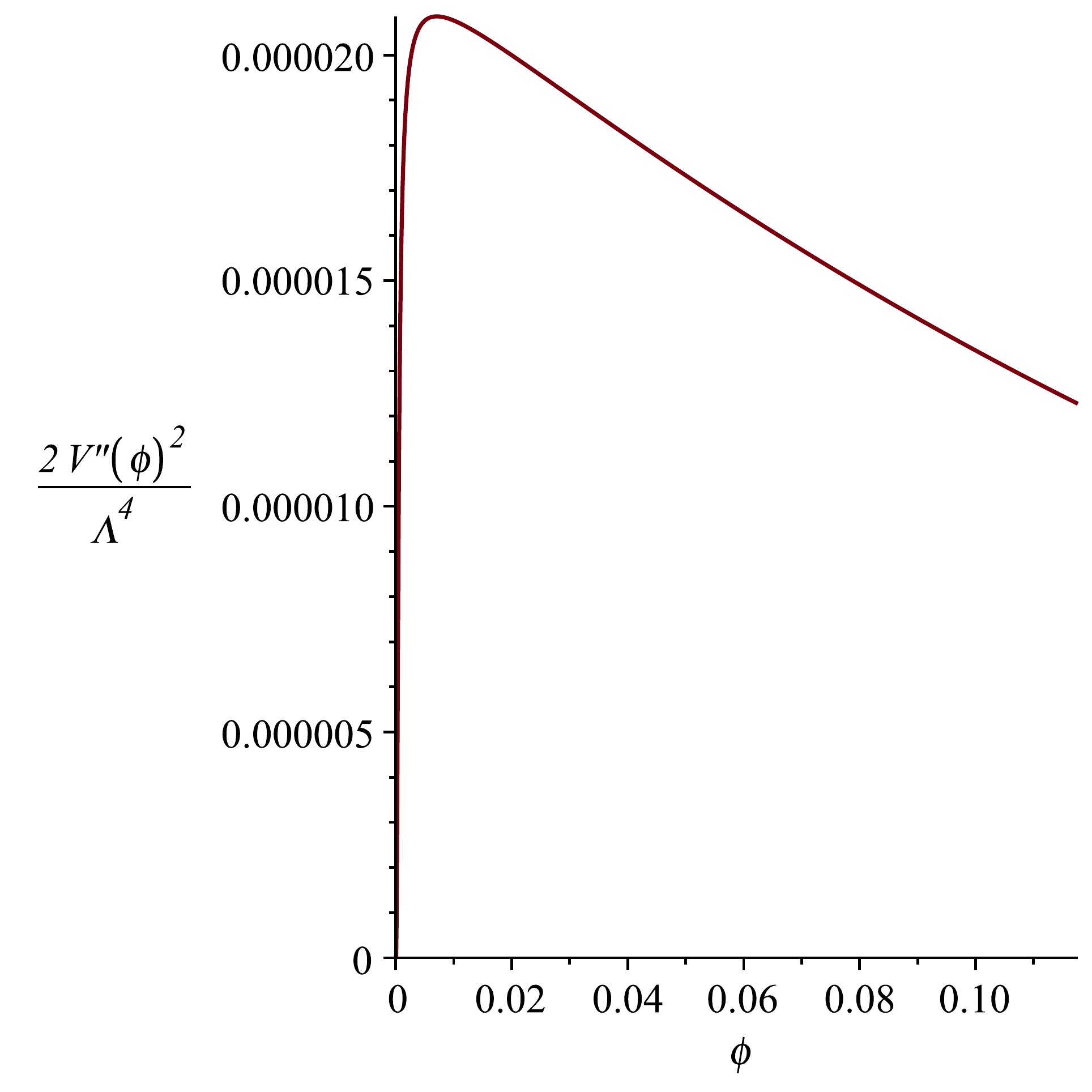}
}
\caption{Here we plot $2V''^2/\Lambda^4$ as function of the {canonical field $\phi$} for larger and smaller ranges of the field for more details. Here $M_P=1,~\xi=1000,~\lambda=0.01$ and $\Lambda=M_P/\xi$.}
\label{fig:arg}
\end{figure}

Masses of effective particles are given by roots of an algebraic equation \cite{Koshelev:2007fi}
\begin{equation}
\Kc_m(m_k^2)=0~\text{ or }~\Fc(m_k^2)=m^2\,.
\label{manysols}
\end{equation}
This has a solution in terms of the Lambert function $W$ as follows\footnote{The Lambert function $W$ solves the equation $xe^x=y$ as $x=W(y)$.}
\begin{equation}
m_k^2=\frac{\Lambda^2}{\sqrt{2}}\sqrt{W(2m^4/\Lambda^4)}\,.
\label{manyW}
\end{equation}
Appearance of infinitely many roots follows from the fact that $W$-function has infinitely many branches.
To see that masses of effective particles $m_k$ are large we need digging into the details of function $W$ \cite{Arefeva:2008zru}. First we compute the argument of $W$ function. In Figure~\ref{fig:arg} the argument in question is plotted and the main observation is that in our case it is bounded roughly by $0.000021$ from above.
The point where the curve touches zero corresponds to the inflection point of the potential.

\begin{figure}[h!]
\center{
\includegraphics[width=7cm]{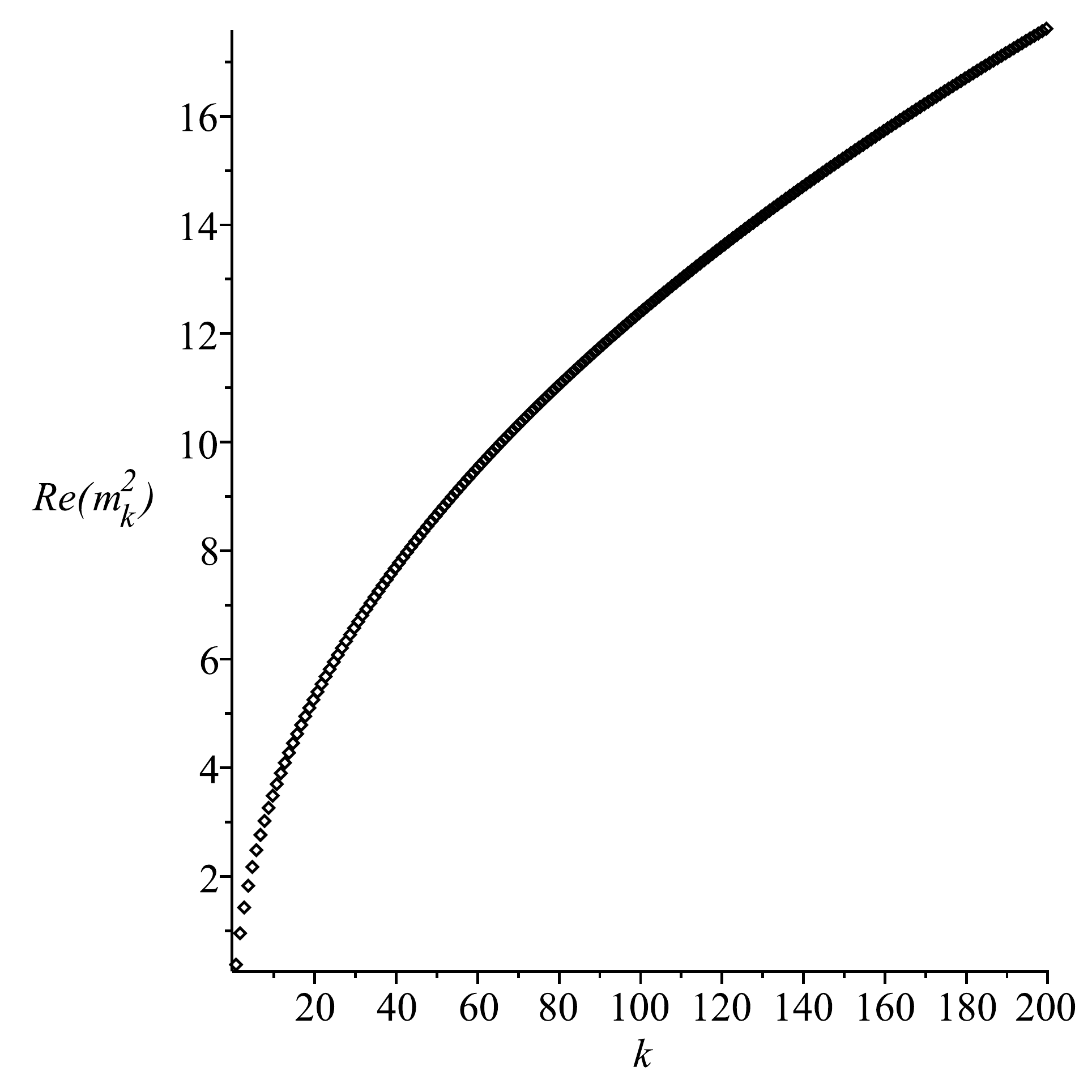}
\includegraphics[width=7cm]{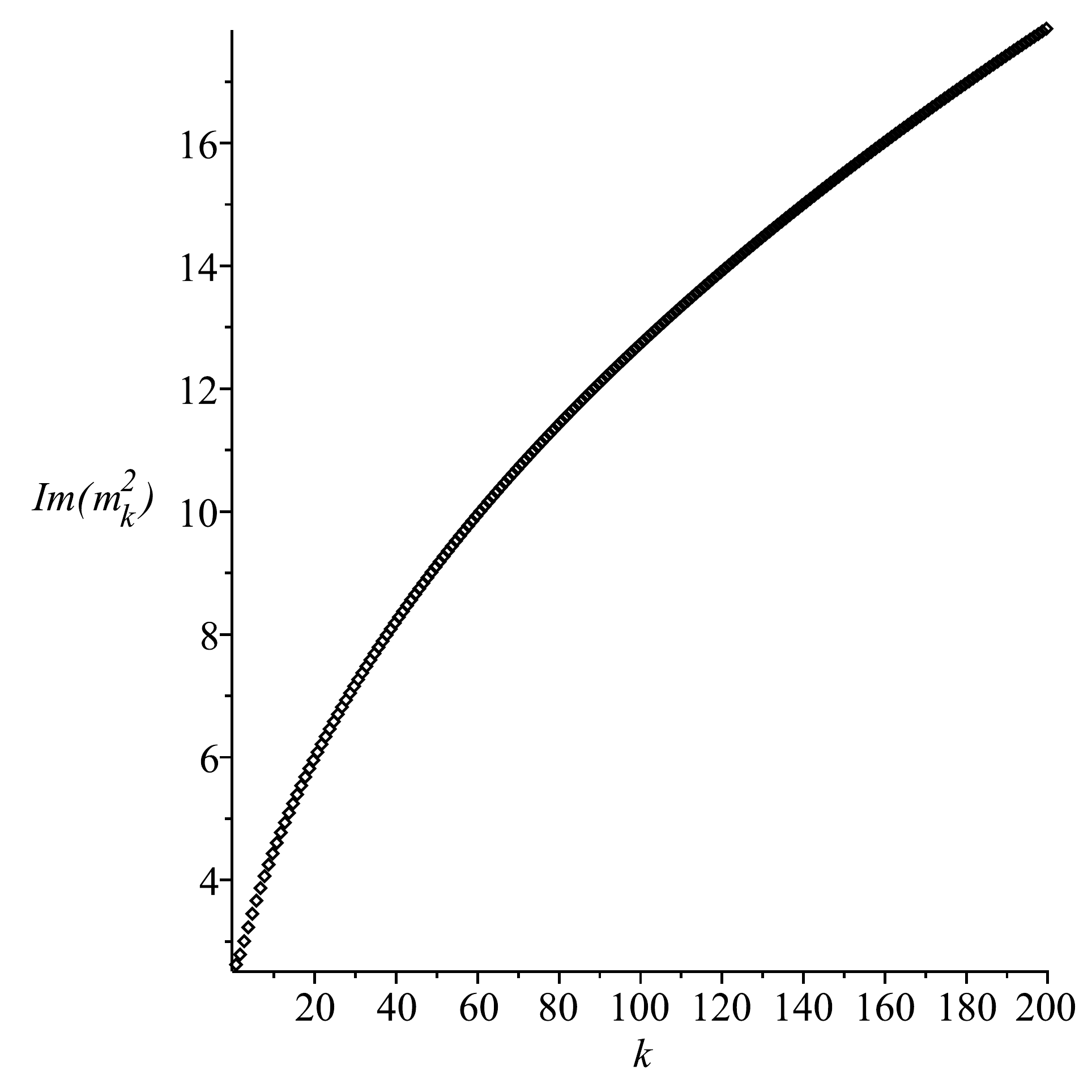}\\
\includegraphics[width=7cm]{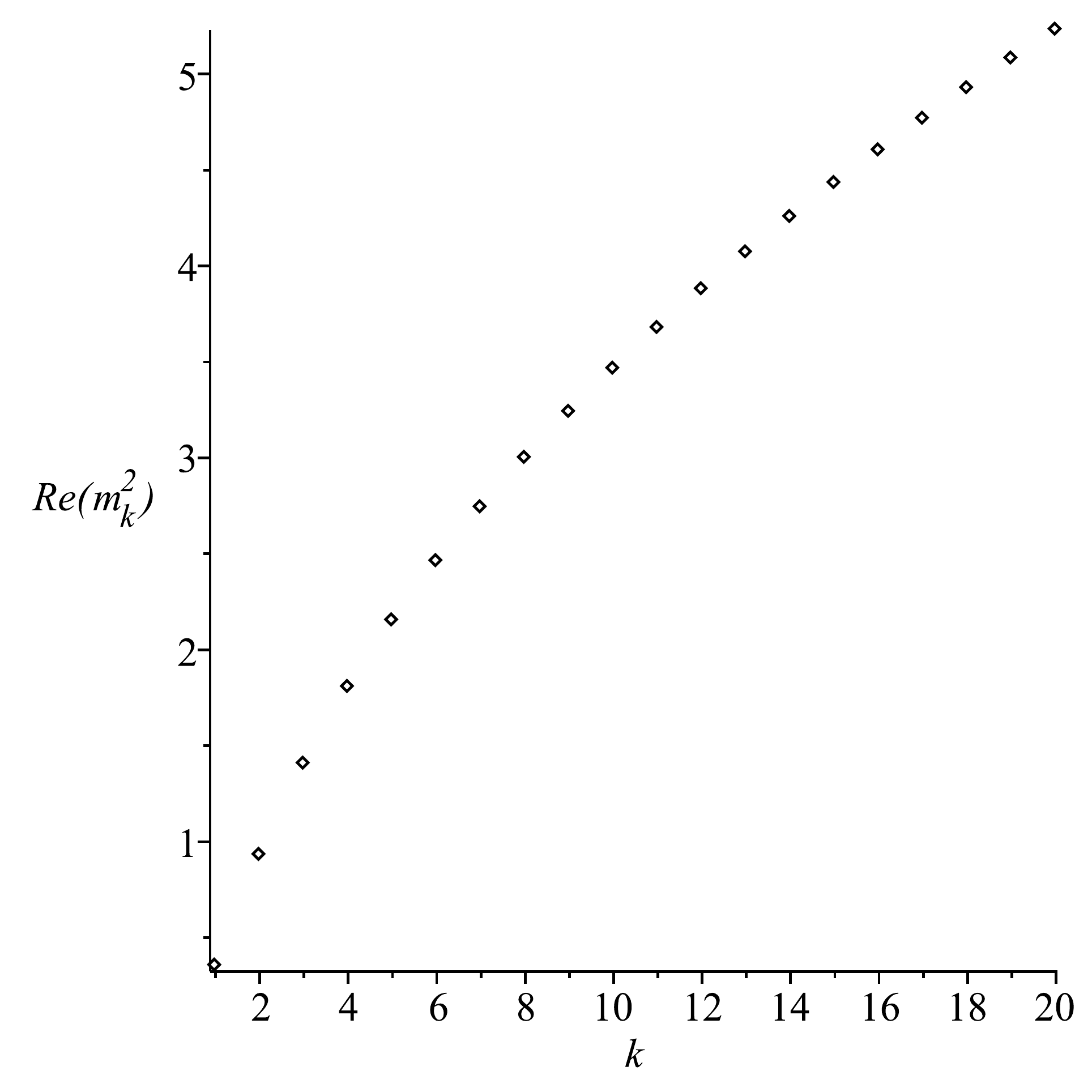}
\includegraphics[width=7cm]{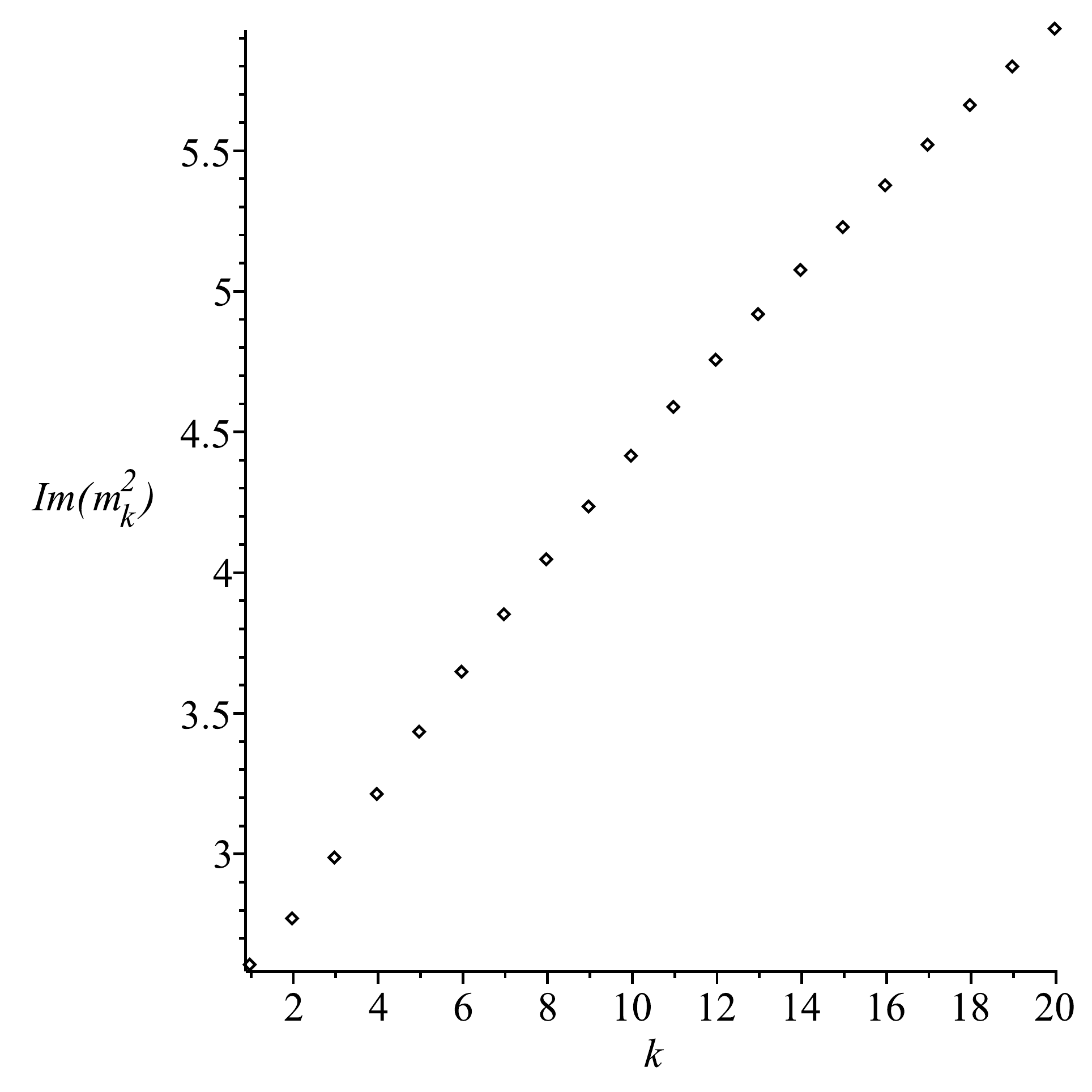}
}
\caption{Real and Imaginary parts of {$m_k^2$ in units of $\Lambda$} calculated using the approximation (\ref{Wklog}). The second row of plots depicts the situation for small branch numbers.}
\label{fig:mk}
\end{figure}

We thus need to study the behavior of the $W$ function for small arguments. Denoting $W_k$ the $k$-s branch of the $W$ function we designate $k=0$ to be the branch which gives a real solution which is in our case unique and corresponds to the mode of interest, the inflaton. All other branches provide the ``would be'' particles with complex masses such that masses are complex conjugate for branches $k$ and $-k$. Moreover, values of the $W_k$ function for $k\neq0$ for small non-negative arguments (which is clearly satisfied in our case as we gain a strictly positive value) can be approximated as
\begin{equation}
W_k(x)\approx\log(x)+2\pi i k-\log(\log(x)+2\pi i k)\,.
\label{Wklog}
\end{equation}
Here $k$ takes discrete integer values as it should be. The complex conjugation upon changing $k\to -k$ is obvious.
In Figure~\ref{fig:mk} we plot the real and imaginary parts of the $\sqrt{W_k(0.000021)/2}$, i.e. the real and imaginary part of $m_k^2$ for the maximum possible value of the argument of the $W$-function for our potential. The smaller the argument, the greater masses will be generated. As such the largest argument provides the worst-case scenario. We see that both real and imaginary parts of $m_k^2$ are greater than unity (with the exception of the real part of $m_1^2$ while still $|m_1^2|>1$) and as such $m_k$ describes somewhat massive field with a mass squared above the non-locality scale.

{In order to come to standard grounds and} to bring an even more intuitive picture we can disband the two unusual complex scalar fields into a pair of real but non-diagonalizable scalar fields as follows \cite{Koshelev:2007fi}:
\begin{equation}
\begin{split}
L=&\frac12\Fc'(m^2)\phi(\square-m^2)\phi+\frac12\Fc'({m^2}^*)\phi^*(\square-{m^2}^*)\phi^*\\
=&\alpha(f\square-f\mu^2+g\nu^2)\alpha-\beta(f\square-f\mu^2+g\nu^2)\beta
-2\alpha(g\square-g\mu^2-f\nu^2)\beta\,.
\end{split}
\label{phictoab}
\end{equation}
Here we use $\Fc'(m^2)=f+ig$, $m^2=\mu^2+i\nu^2$ and $\phi=\alpha+i\beta$ with all quantities used in the last line of the above formula being real. Also we omit the index $k$ for $m_k$ because obviously expressions for each $k$ are similar. In these new notations Figure~\ref{fig:mk} depicts $\mu^2$ and $\nu^2$.

The latter quadratic form is still not that illuminating though because unusual complex fields with complex masses are replaced with real quadratically coupled fields. We cannot clearly designate them to be good or bad but thanks to the normalization freedom which was mentioned earlier we can diagonalize either the derivative or mass quadratic form. The latter seems to be more informative because doing so we will see what are the masses of these ``would be'' excitations. Diagonalization of the mass quadratic form and the further canonical normalization of the kinetic terms of the fields yields
\begin{equation}
L=\frac12\alpha\left(\square-\frac{\mu^4+\nu^4}{\mu^2}\right)\alpha-\frac12\beta\left(\square-\frac{\mu^4+\nu^4}{\mu^2}\right)\beta
-\kappa\alpha\square\beta\,,
\label{massdiag}
\end{equation}
where the factor $1/2$ has been introduced for convenience and
\begin{equation*}
	\kappa=\frac{\nu^2}{\mu^2}\,.%\frac{\sqrt{\mu^4+\nu^4}\sqrt{f^2+g^2}}{f\nu^2+g\mu^2}\,.
\end{equation*}
We have assumed that the factor in front of the kinetic term of the field $\alpha$ which has been absorbed in the fields to achieve the canonical normalization is positive. We have not lost the generality here because flipping this sign is equivalent to exchanging $\alpha\leftrightarrow\beta$.

\begin{figure}[h!]
\center{
\includegraphics[width=7cm]{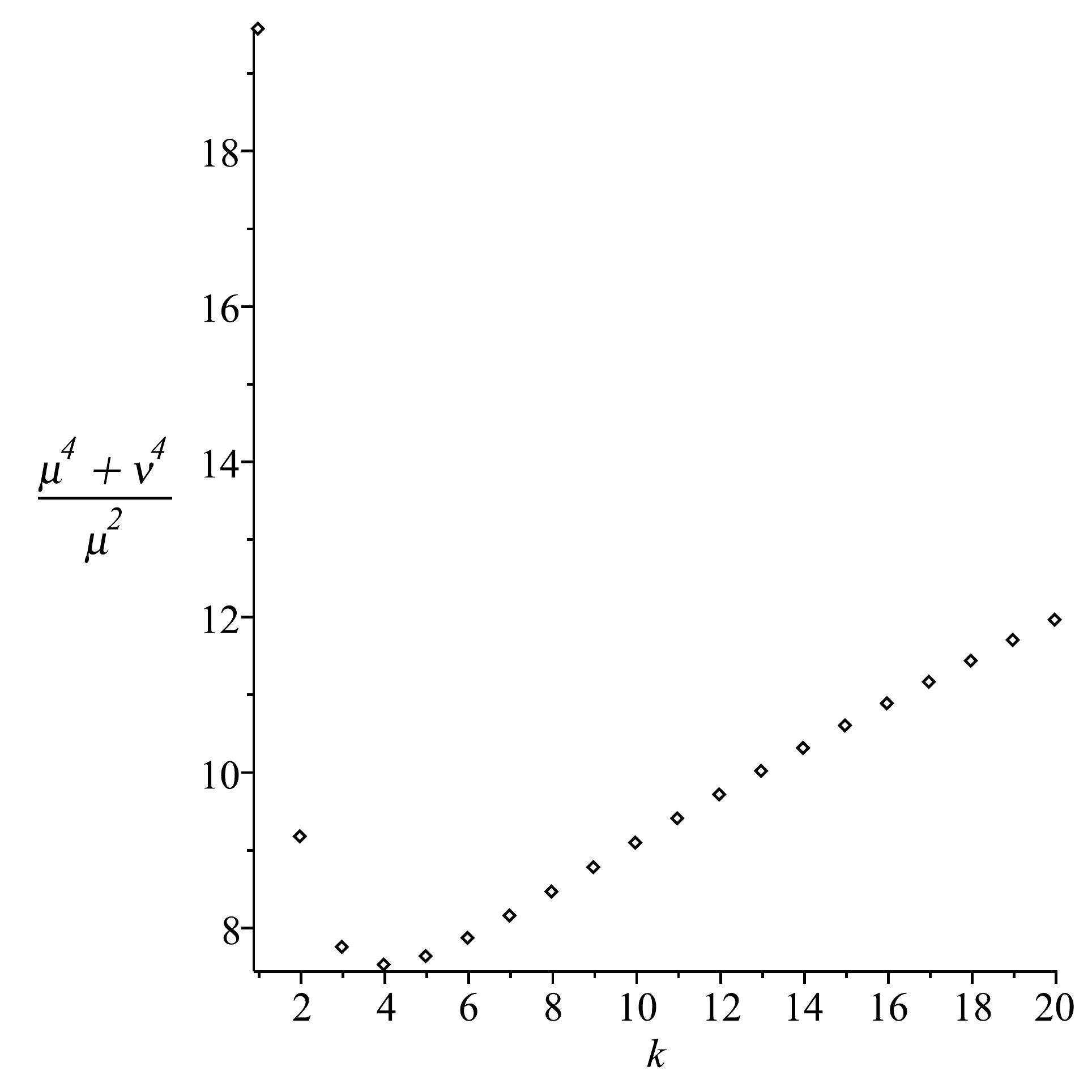}
\includegraphics[width=7cm]{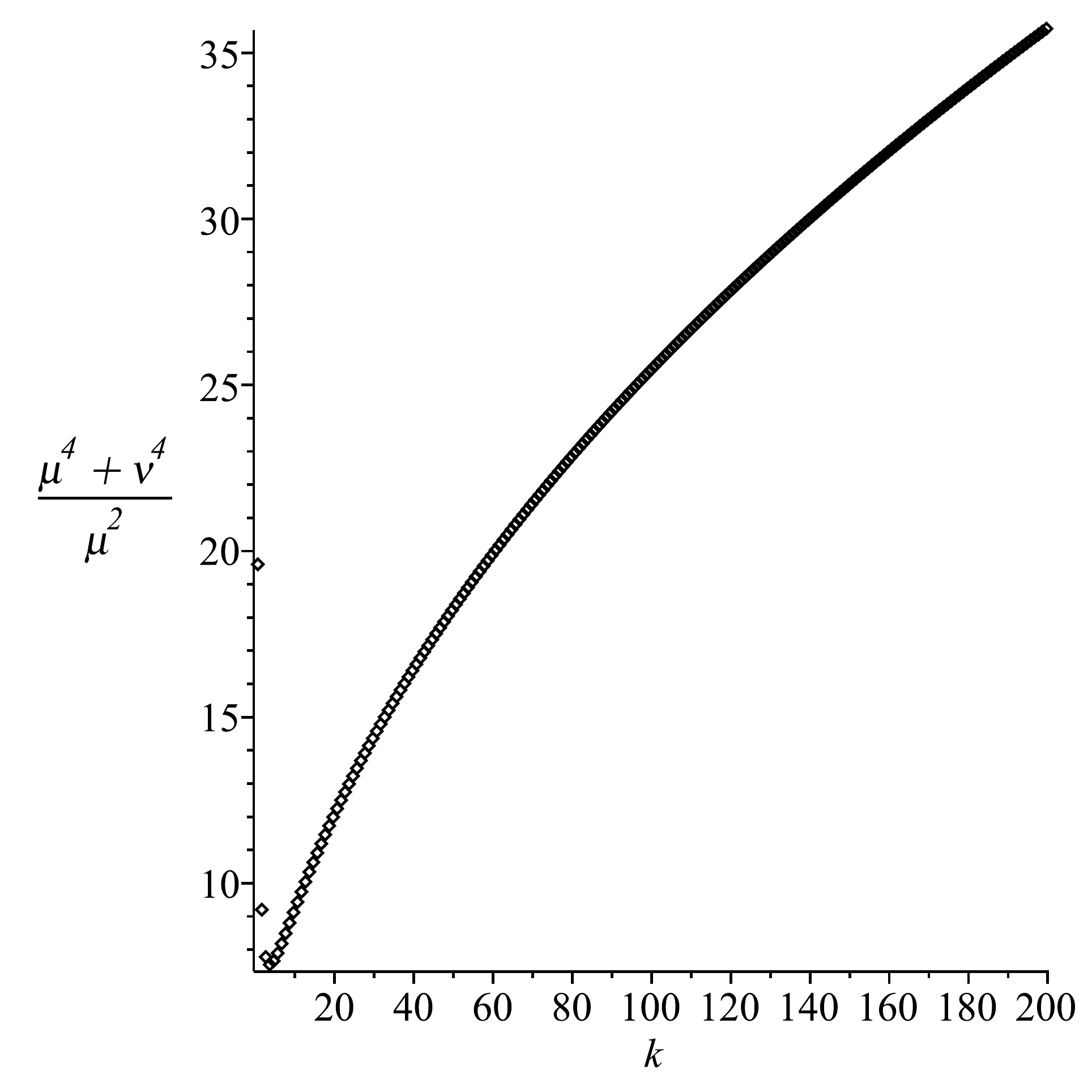}
}
\caption{Mass of the real fields for small and large branch numbers, {in units of $\Lambda$.}}
\label{fig:mreal}
\end{figure}

\begin{figure}[h!]
\center{
\includegraphics[width=7cm]{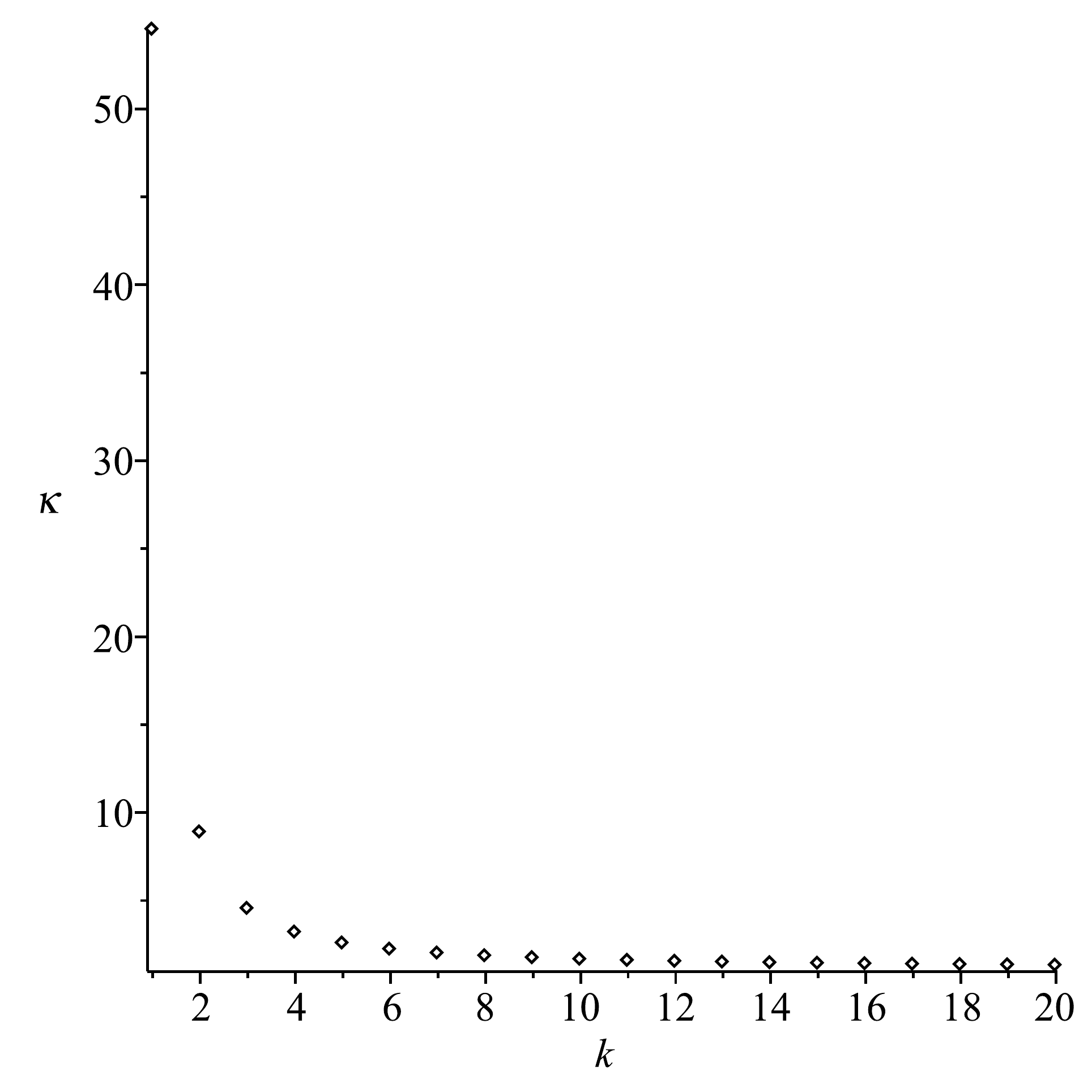}
\includegraphics[width=7cm]{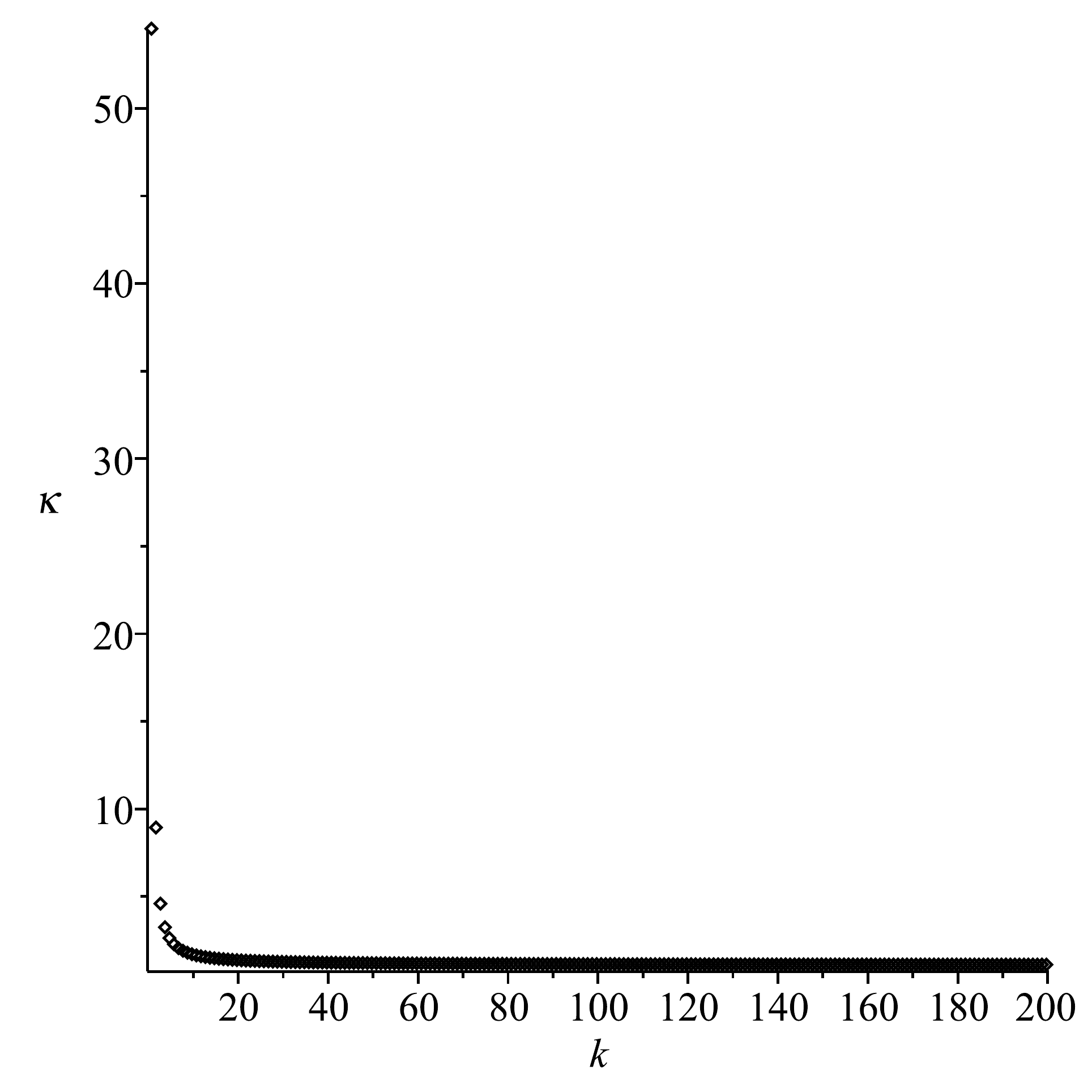}
}
\caption{Mixing coefficient $\kappa$ for small and large branch numbers, {in units of $\Lambda$.}}
\label{fig:mix}
\end{figure}

What happens is that the masses of real fields are very large compared to the cutoff scale. In Figure~\ref{fig:mreal} we see their behavior for small and large values of the branch number. The mixing factor $\kappa$ quickly stabilizes to unity (see Figure~\ref{fig:mix}) and does not affect the interpretation that heavy masses of new effective modes prevent their excitation during evolution characterized by energy scales below the parameter $\Lambda$.

Still we want to see that it is possible to have no growing classical solutions for these new effective modes. This simply boils down to solving the free equations of motion of the form
\begin{equation}
(\square-m_k^2)\phi=0
	\label{nogrow}
\end{equation}
for all modes enumerated by $k$ when the background d'Alembertian operator is evaluated on the de Sitter background. Even thought the equation looks familiar it gets a new twist because $m_k^2$ is complex. Given that the Hubble parameter is denoted as $H$ for our background de Sitter space-time the solution to the latter equation is given by
\begin{equation}
\phi=e^{-\frac32Ht}\left(\alpha J_\rho\left(\frac{ka_0}He^{-Ht}\right)+\beta Y_\rho\left(\frac{ka_0}He^{-Ht}\right)\right)\text{ with }\rho=-\sqrt{\frac94-\frac{m_k^2}{H^2}}
	\label{nogrosol}
\end{equation}
where $J_\rho,Y_\rho$ are Bessel functions of the first and second kind and $a_0$ is the normalization of the scale factor in the metric tensor at $t=0$ and $\alpha,\beta$ are integration constants. Absence of growing solutions means that for large times $t$ which correspond to small arguments of the Bessel functions both branches with coefficients $\alpha$ and $\beta$ at most freeze to constants or have non-growing oscillations. This results in the demand that both functions in the solution grow at most as $e^{3Ht/2}$. Series expansion of Bessel functions with index $\rho$ near the origin tells
\begin{equation*}
J_\rho(x)\sim x^\rho,~Y_\rho\sim x^{-\rho}
\end{equation*}
and it follows from here that we are good to go if ${\rm Re}({\rho})\leq3H/2$. Upon some algebra one can figure out that this corresponds to
\begin{equation}
	({\rm Im} (m_k^2))^2<9H^2{\rm Re}(m_k^2)
	\label{parabola}
\end{equation}

This selects the interior of a parabola-shaped domain on the complex plane. All solutions in (\ref{manysols}) must satisfy this condition. This condition prompts for a careful choice of an entire function in the exponent of the infinite-derivative operator and we see that our simple choice of $\square^2$ does not fulfill the formulated requirement. However, known facts in the complex analysis do not impose any restriction to have such a function and one can try to obtain the desired behavior by combining the Cauchy integral representation for holomorphic functions and the Weierstrass decomposition valid for entire functions \cite{1106.3439}. In particular one can deduce the following sufficient condition on the entire function in the exponent of the infinite derivative operator: the absolute value of this function should grow to infinity only along the positive real ray and be bounded in any other direction.   It is important to mention that the interpretation drawn from eq. (\ref{massdiag}) absolutely holds as long as the ${\rm Re}(m_k^2)\gg\Lambda$.

%%%%%%%%%%%%%%%%%%%%%%%%%%%%%%%%%%%%%%%%%%%%%%%%%%%%%%%%%%%%%%%%%%%%

\subsection{Classical background equations}

{Let us now proceed with constructing a model leading to the early Universe inflation which (up to small corrections) resembles the Higgs inflation model written in the unitary gauge. Alternatively, one can look at this construction as building a UV completion for the inflation driven by the real scalar field with large non-minimal coupling. The only modification required is the non-local kinetic term for the scalar field in the Einstein frame. Here we show that this modification does not spoil inflationary solution in the slow-roll regime.}

Freedman equations for non-local model \eqref{NLlagrangian} look as follows \cite{Koshelev:2010bf}:
\begin{eqnarray}
3 M_P^2 H^2 &=\rho \,,\\
\Fc(\partial_t^2+3H\partial_t) h+ V_h &=0\,.
\end{eqnarray}
If $\Fc(\square)=\Sigma_{k\geq0} f_k\Box^k$ the energy density is
\begin{equation}
\begin{split}
&\rho= \frac{1}{2}\Sigma_{n=1}^{\infty} f_n \Sigma_{l=0}^{n-1}\left[\partial_t(\partial_t^2+3H\partial_t)^l h \partial_t(\partial_t^2+3H\partial_t)^{n-l-1}h+\right. \\
&\left.+(\partial_t^2+3H\partial_t)^l h(\partial_t^2+3H\partial_t)^{n-l} h - \frac{1}{2} h F(\partial_t^2+3H\partial_t) h\right] +V(h)\,.
\end{split}
\end{equation}

Looking for the first correction coming from the higher derivative modification which is $\Fc(\Box)=\Box+\frac{\Box^2}{\Lambda^2}+\dots$ one gets:
\begin{equation}
\rho=\frac{1}{2}\dot{h}+V+\frac{1}{\Lambda^2}\left(\dot{h}\partial_t(\partial_t^2+3H\partial_t)h+((\partial_t^2+3H\partial_t)h)^2\right)\,,
\end{equation}
\begin{equation}
\ddot{h}+3\left(H+\frac{\ddot{H}}{\Lambda}\right)\dot{h}+V_h+\frac{h^{(4)}}{\Lambda^2}+\frac{6 H}{\Lambda^2}h^{(3)}+\frac{6\dot{H}+9 H^2}{\Lambda^2}\ddot{h}=0\,.
\end{equation}
In the slow-roll approximation all higher derivative terms are suppressed even if the non-locality scale is of order of the Hubble scale. Indeed, since $h^{(3)}\sim \eta^2 H^2\dot{h}$ and $h^{(4)}\sim \eta^3 H^3\dot{h}$ ($\eta=V''/(M_P^2 V)$) the leading order slow-roll approximation is not affected by the non-locality. For this reason, the inflationary solution is the same as in the original Higgs inflation.

{We see here that even though one field is nominally present in the Lagrangian, this field has a non-local higher derivative quadratic form leading to higher derivative corrections to any computed values as long as the next orders in the slow-roll approximation are taken into account. This should generically lead to violation of the standard consistency relation for non-Gaussian corrections present in single-field models \cite{Maldacena:2002vr,Creminelli:2004yq,Chen:2006nt}. The expectation is to see this consistency relation altered in higher orders in the slow-roll parameter by values of derivative of the form-factor $\Fc(\square)$ at zero (because the mass of the inflaton field, i.e. the Higgs field, is zero in the inflationary vacuum). Then the picture will resemble the one got for non-Gaussian correction in case of the Starobinsky inflation in the framework of AID gravity \cite{Koshelev:2020foq}.
}

\subsection{Effective potential}

As the next step we explicitly compute the one-loop correction to the inflaton potential in a model with non-local propagator. One-loop correction to the effective action has the following general form:
\begin{equation}
S_1=-\frac{i}{2} \log{{\rm det}\left(\frac{\delta^2 L}{\delta\phi^2}\right)}\,.
\end{equation}
Corresponding correction to action (\ref{lagrangian})can be computed using
\begin{equation}
\label{effective action}
S_1=-\frac{i}{2}\int d^4 x\int \frac{d^4 k_M}{(2\pi)^4}\log\left(\Fc(-k_M^2)+V''(\phi)\right)\,.
\end{equation}
{Here $k_M$ is the physical momentum in Minkowski space.} The effective action is a functional of the only scalar variable $V''(\phi)$ which means that the analytic continuation of the momentum integral from Euclidean to the Minkowski space is straightforward.\footnote{Transition from Minkowskian to Euclidean momenta can be an issue in
non-local theories for the amplitudes which depend on external momenta.
The familiar Wick rotation is not applicable in this case. Here we do
not go deep into details of this consideration and follow the results of
\cite{Pius:2016jsl}. The crux of the new prescription is to perform the internal loop
integration assuming that all momenta are Euclidean and do the analytic
continuation of the external momenta to the Minkowskian signature after
all the internal momenta integration has been carried out using the
Euclidean signature. The paper \cite{Pius:2016jsl} proves that this approach preserves
the unitarity. Moreover, in case of local theories this approach
coincides with the Wick rotation prescription.
} The integral can be computed for Euclidean momentum from the very beginning which would lead to the disappearance of the factor $-i$ in (\ref{effective action}). Here we assume $\Fc(k^2)=k^2 e^{k^4/\Lambda^4}$. Formally, the integral is divergent but it can be rewritten as
\begin{equation}
\begin{split}
L_1(V''(\phi))=\int \frac{d^4 k}{2(2\pi)^4}\log\left(k^2 e^{k^4/\Lambda^4}+V''(\phi)\right)=\\
=\int \frac{d^4 k}{2(2\pi)^4}\log\left(k^2 e^{k^4/\Lambda^4}\right)+\int \frac{d^4 k}{2(2\pi)^4}\log\left(1+\frac{V''(\phi)e^{-k^4/\Lambda^4}}{k^2}\right)\,.
\end{split}
\end{equation}
The first term on the last line is just an infinite contribution to the normalization of the functional integral as it does not depend on the background field. For this reason, we can consider only the last term which is convergent
\begin{equation}
I(V'')=\int \frac{d^4 k}{2(2\pi)^4}\log\left(1+\frac{V''(\phi)e^{-k^4/\Lambda^4}}{k^2}\right)\,.
\end{equation}
If we start with the tree level potential of the form $V=m_0^2\phi^2/2+\lambda \phi^4/4$ we obtain using $V''=3\lambda\phi^2+m_0^2$ a large finite correction to the mass term,
\begin{equation}
V_{eff}=-I(m_0^2)-3\lambda I'(m_0^2)\phi^2+\frac{1}{2}m_0^2\phi^2+\frac{\lambda}{4}\phi^4\,.
\end{equation}
Since $I'(m_0^2)\sim \Lambda^2$ we get a correction to the mass term of order of the non-locality scale. In a local theory corrections of such type correspond to quadratic divergences. In non-local theory on contrary this contribution is finite but it can still be large. If we want to keep the field massless we need to tune the initial mass $m_0$ in such a way that the condition
\begin{equation}
I'(m_0^2)=\frac{m_0^2}{6\lambda}\,,
\end{equation}
is satisfied. It is an algebraic equation on the number $m_0$, thus it can be solved numerically. Upon this tuning the one-loop correction to the initial potential appears to be small.

\begin{figure}[h]
%\begin{minipage}[h]{0.49\linewidth}
\center{\includegraphics[width=\linewidth]{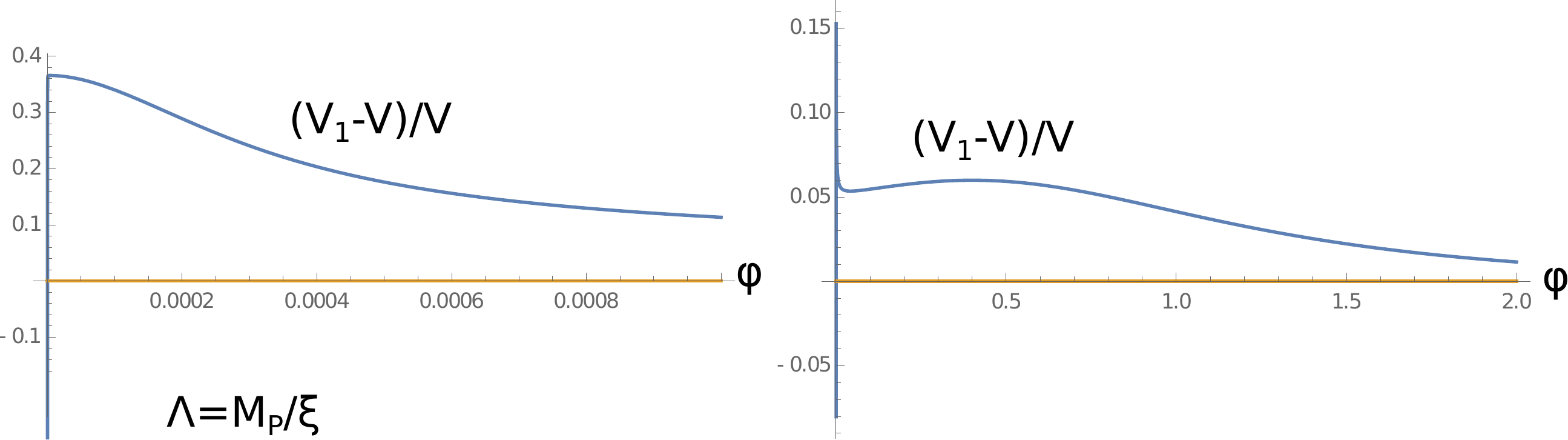} \\ a)}
%\end{minipage}
%\hfill
%\begin{minipage}[h]{0.49\linewidth}
\center{\includegraphics[width=\linewidth]{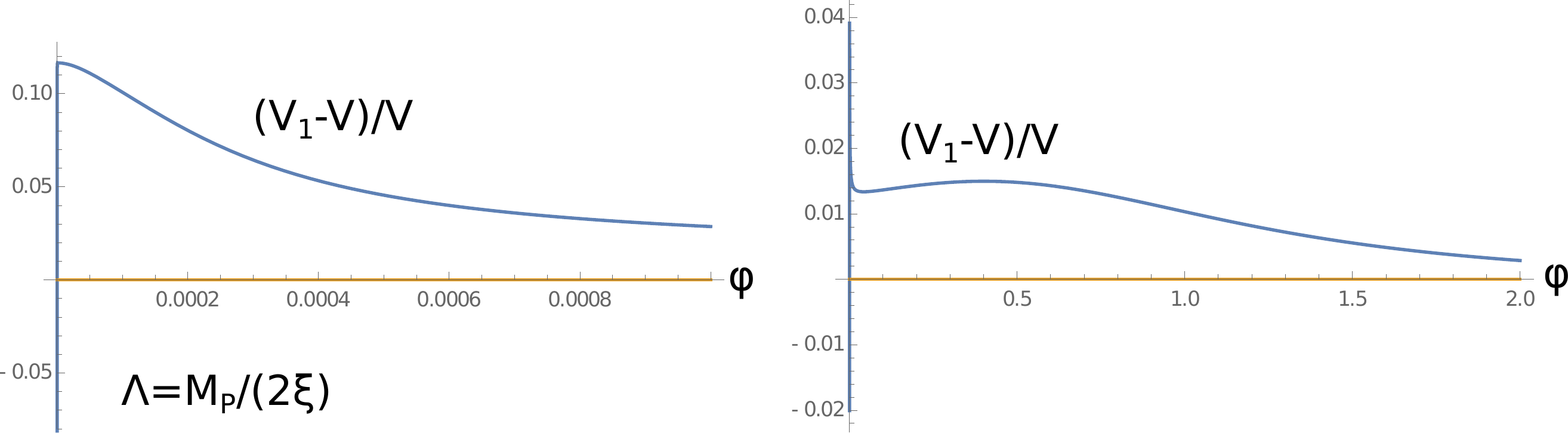} \\ b)}
%\end{minipage}
\caption{Relative deviation $(V_1(\phi)-V(\phi))/V(\phi)$ between the one-loop corrected potential and the classical one for the two choices of non-locality scale.}
\label{ris:image1}
\end{figure}

In a model of inflation with non-minimal coupling (\ref{NLlagrangian}), the potential is more complicated. But the same procedure can still be performed there. Again, tuning of the mass of the field is required. The point is that the simplest way of the addition of a mass term would spoil the flatness of the potential at large field values. Therefore, we should add a term that behaves as a mass at small values of the field and on top of this is suppressed for large fields. Surely, this can be done in an infinite number of ways. Here we present an example that leads to small one-loop corrections to the original (massless) potential. Namely, we take the bare potential of the form:
\begin{equation}
\label{bare potential}
V_0(\phi)=\frac{\lambda M_P^4 \,h(\phi)^4}{4(M_P^2+\xi h(\phi)^2)^2}+\frac{m_0^2\phi^2}{2(1+\xi^2 h(\phi)^4)}\,.
\end{equation}
We can show numerically that given the non-locality scale is smaller than $M_P/\xi$, the one-loop corrected potential $V_1(\phi)=V_0(\phi)-I(V''(\phi))$ is very close to the original one $V(\phi)$ (\ref{NLlagrangian}). The result is presented in Figure~\ref{ris:image1}.

\begin{figure}[h!]
\center{\includegraphics[width=0.5\textwidth]{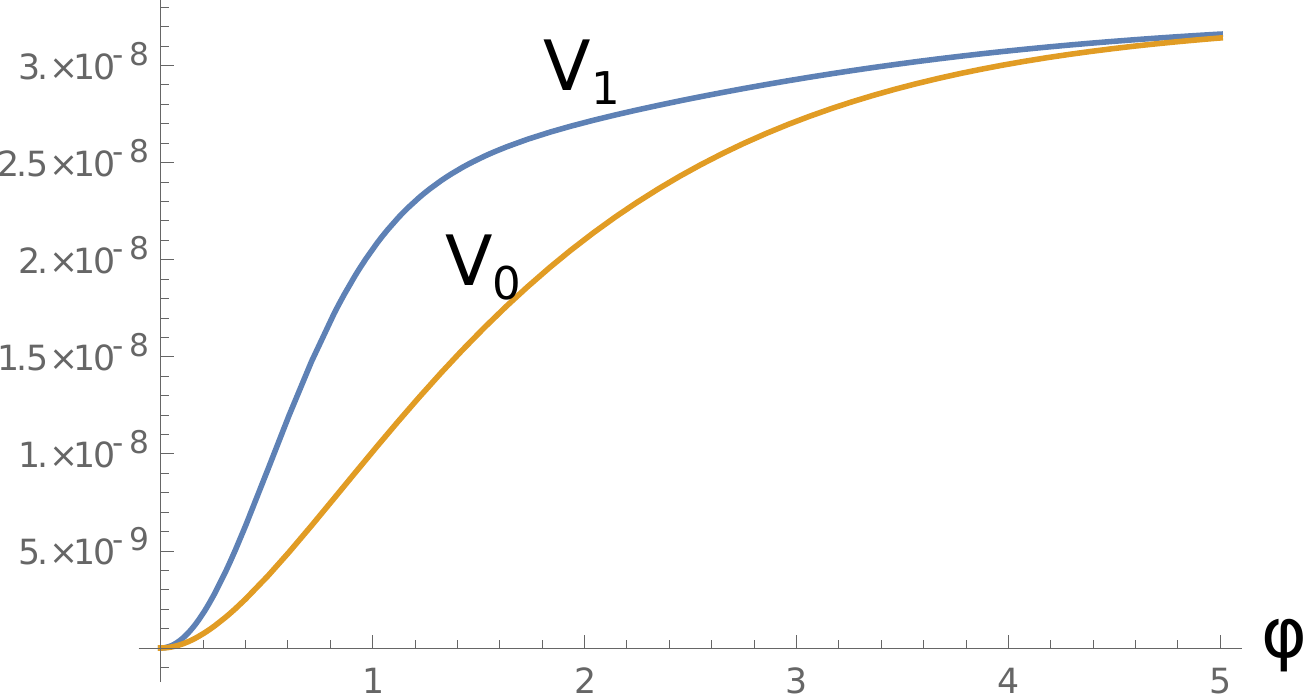}}
\caption{Original and corrected potentials for $\Lambda=5 M_P/\xi$. For $\Lambda=M_P/\xi$ they are practically indistinguishable.}
\label{ris:image2}
\end{figure}

Thus, we have obtained that after tuning of the mass term in the potential the one-loop correction can be made small. This is possible only if the non-locality scale is smaller than the scale $M_P/\xi$. Otherwise, for higher non-locality scales the one-loop correction becomes large, see Figure~\ref{ris:image2}. In this situation, the one-loop result can not be trusted anymore providing a sign of entering the strong coupling regime. Still in the case $\Lambda\lesssim M_P/\xi$ we can not immediately conclude that the model can be treated perturbatively because accurate computations of loop scattering amplitudes is required. We leave this issue for future projects.

\section{Non-locality in the Standard model}

Although we have obtained that a single field model with non-minimal coupling to gravity can be UV completed by an appropriate non-local form-factor one would find that an implementation of this mechanism in the Standard Model is not straightforward. The Standard Model Higgs boson is a complex doublet charged under $SU(2)$ group. This makes the introduction of a form-factor more tricky. {We suggest below a model in which the non-locality is introduced only for the radial Higgs component and show that it can restore the tree-level unitarity at least for $2\rightarrow 2$ scattering amplitudes. }

\subsection{Restoring gauge invariance}

Action (\ref{EFlagrangian}) for non-minimally coupled scalar actually resembles those for the Standard Model Higgs in a unitary gauge $H=(0,(r+v)\sqrt{2})^T$ if the gauge bosons are switched off. Now we add the gauge bosons and restore the full action invariant with respect to the SM gauge transformations.

A covariant derivative in action (\ref{action}) in the unitary gauge can be written as,
\begin{equation}
|D_{\mu} r|^2=\frac{1}{2}(\partial_{\mu} r)^2+\frac{g^2 (v+r)^2}{4}W_{\mu}^+W_{\mu}^-+\frac{(g^2+g'^2)(v+r)^2}{8}Z_{\mu}Z_{\mu}\,.
\end{equation}
Here $v$ is the Higgs VEV and $g,~g'$ are the $SU(2)$ and $U(1)$ SM gauge couplings respectively.
Note that $(\partial_{\mu}r)^2$ can be written in a covariant form
\begin{equation}
(\partial_{\mu}r)^2=\frac{(\partial_{\mu}(H^{\dagger} H))^2}{4\,H^{\dagger} H}\,.
\end{equation}
Therefore, the part of the Lagrangian for Higgs inflation which does not contain the radial mode is
\begin{equation}
L_1=-M_P^2\frac{|D_{\mu} H|^2-\frac{(\partial_{\mu}(H^{\dagger} H))^2}{8\,H^{\dagger} H}}{M_P^2+2\xi H^{\dagger}H}\,.
\end{equation}
Further, the covariant form of the action for the radial mode is
\begin{equation}
L_2=-\frac{(\partial_{\mu}(H^{\dagger} H))^2}{8\,H^{\dagger} H(M_P^2+2\xi H^{\dagger} H)}-3\xi^2 \frac{(\partial_{\mu}(H^{\dagger} H))^2}{(M_P^2+2\xi H^{\dagger} H)^2}-V(H^{\dagger}H)\,.
\end{equation}
In the sequel we propose to not introduce non-locality to the part $L_1$ of the action because extra gauge-covariant derivatives will lead to infinitely many new couplings between SM gauge bosons. In such a complicated framework it would be hard to say whether this leads to a self-consistent finite theory or not. We leave this question open for now.

Hence we suggest modifying only the sector of the radial Higgs mode. However, one should be careful with the introduction of a higher derivative form-factor because it can lead to the fast growth of tree-level amplitudes at energies higher than the cutoff scale. For example, the following Lagrangian
\begin{equation}
L=\frac{1}{2}\phi \Box f(\Box)(\phi+a \phi^2)\,,
\end{equation}
leads to exchange diagrams which grow rapidly at large momenta, since we would choose the sign in front of the d'Alembertian operator such that loop integrals converge.

The way to proceed is to introduce non-locality in the same way as it was done in Section~2 and this can be safely accomplished. Namely, we canonically normalize the kinetic term and modify the canonical propagator. This would not lead to pathologies in the sector of the radial Higgs mode. Still, we are going to show in the next Section that scattering of longitudinal modes of vector bosons would grow with momenta starting from $M_P/\xi$.

\subsection{Issue of tree-level unitarity}

Besides the strong-coupling in the scattering amplitudes of the radial Higgs mode, the Higgs inflation model suffers from the problem of the tree-level unitarity in the sector of gauge bosons \cite{Bezrukov:2010jz,Bezrukov:2014ipa}. Let us show the origin of these problems in a somewhat simpler model of a $U(1)$ field. This model resembles the same issue which appears in the non-Abelian case. The scattering amplitudes of longitudinal polarizations of gauge bosons are known to grow as $\xi p/M_P$ violates unitarity at $M_P/\xi$.

\begin{figure}[h!]
\center{\includegraphics[width=\textwidth]{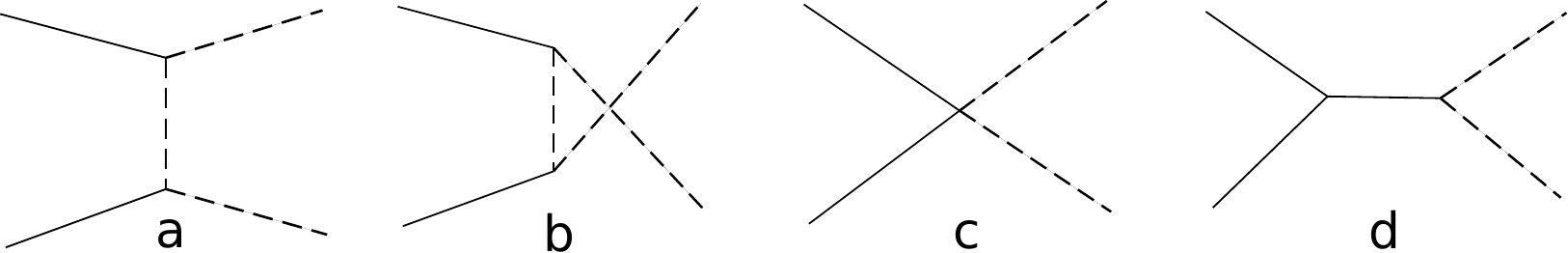}}
\caption{\footnotesize{Diagrams for tree-level scattering $hh\rightarrow\theta\theta$.}}
\label{ris:diagrams}
\end{figure}

We work in the so called $R\xi$ gauge \cite{Langacker:2010zza} where the longitudinal part of the Abelian vector field is kept in a phase of the Higgs, $H=h e^{i\theta}/\sqrt{2}$.\footnote{The longitudinal part $z$ can be written as $H=(h_0+i z)/\sqrt{2}$. Due to the equivalence theorem \cite{Langacker:2010zza} we study scatterings of the Goldstone modes instead of the scatterings of gauge bosons} In this case, in terms of the canonical variables, our Lagrangian has the form,
\begin{equation}
L=-\frac{1}{2} (\partial_{\mu} h)^2 - \frac{1}{2} G(h) (\partial_{\mu} \theta)^2 - V(h)\,.
\end{equation}
In the Standard Model $G(h)=h^2$. In general, $G(h)=G(v)+G'(v) h+G''(h) h^2/2+...$. We are interested only in the amplitudes which grow with momentum. $hh\rightarrow hh$ amplitude does not grow at tree level since there are no derivative couplings. But $hh\rightarrow \theta\theta$ amplitude will grow unless $G(h)=h^2$. This growing is coming from three diagrams, a, b, c in Figure~\ref{ris:diagrams}, and it has a form
\begin{equation}
{\cal M}\sim s\left(\frac{2 G''(v)}{G(v)}-\left(\frac{G'(v)}{G(v)}\right)^2\right)\,.
\end{equation}
The diagram $d$ in Figure~\ref{ris:diagrams} does not lead to an amplitude which grows with momentum.
In the case of Higgs inflation, approximately finding potential (\ref{potential}) we get
\begin{equation}
\phi(h)\approx h-\frac{2 \xi^2}{M_P^2}h^3\,.
\end{equation}
This leads to
\begin{equation}
G(h)=h^2(1-\frac{4\xi^2}{M_P^2}h^2), \quad {\cal M}\sim \frac{\xi^2 s}{M_P^2}\,.
\end{equation}

Non-locality, as it is introduced in Section~2 in the radial sector, can not treat this behavior since no changes are incurred for these particular diagrams. However, the problem can be solved if a non-local form-factor is added in a more complicated way. To see this we exploit the freedom in splitting quadratic and interaction parts in the Lagrangian. Namely, if we define a new field variable
\begin{equation}
y(h)=\frac{h M_P}{\sqrt{M_P^2+\xi h^2}}\,,
\end{equation}
this would bring the coupling to the safe form $y^2(\partial_{\mu}\theta)^2$. Then, the kinetic term can be split as follows
\begin{equation}
\label{Ly}
L=-\frac{1}{2}(\partial_{\mu} y)^2-\frac{\xi y^2(\partial_{\mu}y)^2((2+6\xi)M_P^2-\xi y^2)}{2(M_P^2-\xi y^2)^2}- V(y)\,.
\end{equation}
If we designate $y$ to be our canonical field variable and treat the second term as an interaction than in this case we get growing amplitude $yy\rightarrow \theta\theta$ from the $s$-channel diagram with $y$-exchange (diagram d in Figure~\ref{ris:diagrams}). {Note that this diagram did not give growing with momentum contribution in the original variable $h$.} Besides this diagram, all other diagrams which impact this scattering process (a, b, c) do not grow. {Thus, in these variables, we narrowed the problem of tree unitarity breaking to the sector of radial the Higgs mode.} This means that once we introduce a non-local form-factor in the first term in (\ref{Ly}) we arrive at the exponential suppression of this diagram (due to non-local propagator) which solves the problem under discussion.

Notice that the interaction term in (\ref{Ly}) does not spoil the nice features of the model such as convergent Higgs loops and radial Higgs scattering amplitudes. Tree level scattering amplitude $hh\rightarrow hh$ does not grow due to the crossing symmetry. Indeed, the matrix element would scale as $s+t+u$ in Mandelstam variables and the latter combination is constant on-shell. {However, scattering amplitudes of many particles can still suffer from the unitarity breaking at scales lower than Planck mass. In principle, this can be cured by appropriate non-local modification of the second term of (\ref{Ly}). In general, such modifications are not expected to spoil the background solution for Higgs inflation in the slow-roll regime since they modify only terms with derivatives which are suppressed in the slow-roll approximation.}

Eventually, let us present a covariant form of a non-local Higgs Lagrangian which provides Higgs inflation and is safe from the strong coupling problem at tree level {which originally appears at the scale $M_P/\xi$.}
\begin{equation}
\begin{split}
L&=\frac{1}{2}y(\sqrt{H^{\dagger}H})\Fc(\Box)y(\sqrt{H^{\dagger}H})\\
&-\frac{\xi y(\sqrt{H^{\dagger}H})^2(\partial_{\mu}y(\sqrt{H^{\dagger}H}))^2((2+6\xi)M_P^2-\xi y(\sqrt{H^{\dagger}H})^2)}{2(M_P^2-\xi y(\sqrt{H^{\dagger}H})^2)^2}\\
&-M_P^2\frac{|D_{\mu} H|^2-(\partial_{\mu}(H^{\dagger} H))^2/(8\,H^{\dagger} H)}{M_P^2+2\xi H^{\dagger}H}- V(H^{\dagger}H)\,.
\end{split}
\end{equation}
Here $\Fc(\Box)=(\Box-m_H^2)e^{\sigma(\Box)}$ with an entire function $\sigma(\square)$ in the exponent and the potential is
\begin{equation}
V(H^{\dagger}H)=\frac{\lambda(H^{\dagger}H-v^2)^2}{4(M_P^2+2\xi H^{\dagger}H)}-\frac{1}{2}m_H^2 y(\sqrt{H^{\dagger}H})^2\,.
\end{equation}
The last term is written because the Higgs mass term is already present in $\Fc(\Box)$.

\subsection{Naturalness of Higgs mass}

Here we discuss the hierarchy problem in the Standard Model related to the small mass given by Higgs VEV compared to both inflation and Planck scales. In general, the tiny Higgs mass can be affected by yet unknown high energy physics. If there are new heavy particles coupled to the Higgs, loop corrections to Higgs propagator would be large, starting from this new energy scale. Thus, fine-tuning of constants at high energies would be required in order to keep the Higgs mass small at low energy. If there are no new particles between the electroweak and Planck scales than one could hope that the Higgs mass can be kept small without extra fine-tuning. In the considered framework, we do not add extra particles. Instead of this, we modify the Higgs propagator at high energies.

\begin{figure}[h!]
\center{\includegraphics[width=0.25\textwidth]{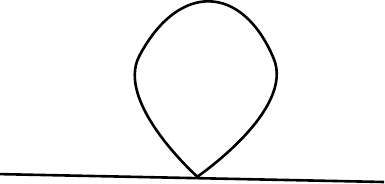}}
\caption{\footnotesize{Loop correction to the Higgs propagator.}}
\label{ris:tadpole}
\end{figure}

In order to address the influence of non-locality on the naturalness issue, we start with the computation of the first loop corrections to the propagator. We skip the diagrams that lead to the corrections proportional to Higgs VEV and start with the diagram in Figure~\ref{ris:tadpole} which is quadratically divergent in the canonical local Standard Model.

Let us compute explicitly the diagram in Figure~\ref{ris:tadpole} coming from the interaction $\lambda h^4/4$ and $\sigma(\Box)=\Box^2/\Lambda^2$. The corresponding expression is given by
\begin{equation}
{\cal M}=\frac{3 i\lambda}{(2\pi)^4}\int\frac{d^4 k e^{-k^4/\Lambda^4}}{k^2-m_H^2+i\varepsilon}\,.
\end{equation}
We follow  \cite{Pius:2016jsl} in defining the relation between Minkowskian and Euclidean
amplitudes. The prescribed definition is briefly outlined in our
footnote 3 and in short it states that performing the analytic
continuation from Euclidean external momenta to physical Minkowski ones
would provide unitary amplitudes. Technically at least for this diagram
we have a very simple case because the answer does not depend on the
external momentum. We thus obtain:
\begin{equation}
{\cal M}=\frac{3 \lambda}{8\pi^2}\int_0^{\infty}\frac{k^3 d k e^{-k^4/\Lambda^4}}{k^2+m_H^2}= \frac{3 \lambda}{8\pi^2}\left(\frac{\sqrt{\pi}}{4}\Lambda^2+\frac{m_H^2}{4}e^{-m_H^4/\Lambda^4}\left(-\pi {\rm Erfi}(m_H^2/\Lambda^2)+{\rm Ei}(m_H^4/\Lambda^4)\right) \right)\,.
\end{equation}
The first term proportional to $\Lambda^2$ reflects the quadratic divergence of the same diagram in a local theory. The performed procedure resembles the higher derivative regularization \cite{Faddeev:1980be} where, first, an infinite tower of derivatives is used, and, second, the regularization parameter is a part of the original system and thus is kept finite, though large, after all the computations are finished.
We note here again that the standard Wick rotation is not directly
applicable in the case of infinite powers of momenta but the result
clearly reduces to the one of a local theory ones the limit
$\Lambda\to\infty$ is performed in which case one can use the Wick
rotation mechanism. This maintains the consistency between the procedure
constructed in \cite{Pius:2016jsl} for non-local theories and the Wick rotation valid
in local theories as long as corresponding parameter limits are
considered. In particular this implies that we should use the same
prescription for the mass pole as in a local theory.

Despite the efforts, however, some tuning of parameters would be needed because the term quadratic in $\Lambda$ would lead to a large correction to the Higgs mass. In order to keep the Higgs mass small, we need its finite renormalization. Namely, if we set the bare mass to be
\begin{equation}
m_{0H}^2=m_H^2-\frac{3\lambda\sqrt{\pi}}{32\pi^2}\Lambda^2\,,
\end{equation}
than at one-loop level we get the SM value of Higgs mass. The price we pay is the fine-tuning of parameters at the scale on non-locality $\Lambda$. This is the one-loop correction which will get modified by higher loops
while further corrections are suppressed by powers of $\lambda$ in full
analogy with the local case.

{This outcome is easy to understand in our approach since we do modify only the behavior of the propagator at large momenta, larger than the non-locality scale while keeping it intact at lower scales. Technically the $\Lambda^2$ contribution has come from the part of integration from zero to $\Lambda$ and this is not affected by the proposed higher derivative modifications in any way. Even more, it seems unfeasible to achieve smallness of such a contribution by adjusting the form-factor because mathematically possible modifications will most likely make the function in the exponent of the form-factor non-entire which in turn will lead to the problem of new excitations, probably ghosts.}

Moreover, it is known to be difficult overall to avoid this kind of fine-tuning.
The most likely scenario if that the Higgs mass can be naturally small only in presence of new symmetries, like super-symmetry \cite{Bardeen:1995kv}, scale invariance without new heavy particles \cite{Shaposhnikov:2008xi} or if for example it is generated non-perturbatively \cite{Shaposhnikov:2018jag}. Finding a symmetry eliminating large quadratic corrections keeping the presented in this paper non-local framework that treats the {renormalizability} issue could be an interesting open direction to study.

\section{Conclusions and discussion}

The main game-changer in the presented consideration compared to previous studies of the Higgs inflation model is the analytic infinite derivative (AID) modification of the scalar field propagator. Namely, an exponent of an entire function of the covariant d'Alembertian is introduced in the kinetic term. Having deep motivations from different perspectives including interacting string theories, such a modification of the propagator will incur major consequences for any model in which it is considered. One of the most important is the ultimate suppression of the loop integrals. To have the things computable we specialize to an exact form of the {non-local form-factor} $\exp(\square^2/\Lambda^4)$. Surely any polynomial with an appropriate sign would work but to keep things tractable one would stick to a monomial. Moreover, a would-be simpler choice of just $\exp(-\square/\Lambda^2)$ does not satisfy our needs as it makes tree-level scattering amplitudes exponentially growing. The important piece here is the mass scale parameter $\Lambda$ which can be thought as an effective cut-off. It is not fixed per se but we, first, would have it above the inflation scale $H$ so that the inflationary dynamics is preserved as in the original local model and, second, it is reasonable to have it of order $\Lambda\sim M_P/\xi$ so that the UV completion by the exponential suppression works at the cut-off scale $M_P/\xi$ of the local model.
Section~2 has all the corresponding analysis which shows that the desired behavior of the model can be achieved. In particular, we have demonstrated explicitly by using the slow-roll approximation that the inflationary dynamics is preserved.

Moreover, we have shown two important things related to the AID model modification. First, we have computed masses of effective heavy excitation which emerge as long as the model is not in a vacuum and have shown that corresponding masses can be easily made way heavier than the non-locality scale which is in turn higher than the Hubble scale. Also upon some adjustments one can achieve that no classical growing solutions are present for these modes. This keeps the setup safe from the influence of extra modes at least at linear order. As a general expectation, we assume those effective excitations may generate non-Gaussian corrections measurably different from the original local setup \cite{Koshelev:2020foq}. A corresponding analysis in this regard would be a nice forthcoming study. Second, we have computed numerically the one-loop effective potential and have observed that it introduces small corrections to the original potential as long as a finite mass renormalization is performed. The latter can be taken as a parameter tuning {after which the loop correction to the original potential remains small}.

It seems natural to expect that the reheating phase will get changed upon introducing the higher derivative modification to the propagator. On the other hand, the fact that the masses off effective heavy excitations are always above the non-locality scale makes us thinking that incurred changes will preserve the main features of the reheating phase keeping the main predictions modified mildly compared to the local model. It is yet one more open question to be investigated soon.

In Section~3 we have shown that the AID modification of the propagator can be adopted in the full SM Higgs inflation setup without problems. Even better, we outline the ideas on how to introduce further modification in the spirit of AID form-factors such that the tree-level unitarity in the sector of gauge bosons is improved. However, the hierarchy problem or the problem of the Higgs mass naturalness still seeks for a better resolution. In our modification, we have achieved that a finite mass term tuning is required which is in a sense better than a quadratic divergence in the original local model. However, finding a symmetry that would preserve quadratic corrections small should be definitely addressed in future studies.

{We can further contemplate that the construction presented in this work for the Higgs inflation model can be merged as a part of a unique non-local UV-complete theory that features AID gravity. Despite a number of open questions, the AID gravity seems to be a good candidate for the unitary and renormalizable description of gravity at very high energies \cite{Biswas:2011ar,Koshelev:2017ebj,Koshelev:2020xby}.}

\section*{Acknowledgements}

Authors would like to thank I.~Aref'eva, F.~Bezrukov, and M.~Shaposhnikov for illuminating and inspiring discussions and D.~Gorbunov and V.~Rubakov for sharp and very interesting questions. AK is supported by FCT Portugal investigator project IF/01607/2015. AT is supported by Foundation for the Advancement of Theoretical Physics and Mathematics BASIS grant. The part of work of AT related to the construction of real Standard Model Higgs with non-local propagator was supported by the Russian Science Foundation grant 19-12-00393.

%\bibliographystyle{utphys}
%\bibliography{anya}

\begin{thebibliography}{10}

\bibitem{Bezrukov:2007ep}
F.~L. Bezrukov and M.~Shaposhnikov, ``{The Standard Model Higgs boson as the
  inflaton},'' \href{http://dx.doi.org/10.1016/j.physletb.2007.11.072}{{\em
  Phys. Lett. B} {\bfseries 659} (2008) 703--706},
  \href{http://arxiv.org/abs/0710.3755}{{\ttfamily arXiv:0710.3755 [hep-th]}}.

\bibitem{Akrami:2018odb}
{\bfseries Planck} Collaboration, Y.~Akrami {\em et~al.}, ``{Planck 2018
  results. X. Constraints on inflation},''
  \href{http://arxiv.org/abs/1807.06211}{{\ttfamily arXiv:1807.06211
  [astro-ph.CO]}}.

\bibitem{Burgess:2009ea}
C.~Burgess, H.~M. Lee, and M.~Trott, ``{Power-counting and the Validity of the
  Classical Approximation During Inflation},''
  \href{http://dx.doi.org/10.1088/1126-6708/2009/09/103}{{\em JHEP} {\bfseries
  09} (2009) 103}, \href{http://arxiv.org/abs/0902.4465}{{\ttfamily
  arXiv:0902.4465 [hep-ph]}}.

\bibitem{Barbon:2009ya}
J.~Barbon and J.~Espinosa, ``{On the Naturalness of Higgs Inflation},''
  \href{http://dx.doi.org/10.1103/PhysRevD.79.081302}{{\em Phys. Rev. D}
  {\bfseries 79} (2009) 081302},
  \href{http://arxiv.org/abs/0903.0355}{{\ttfamily arXiv:0903.0355 [hep-ph]}}.

\bibitem{Bezrukov:2010jz}
F.~Bezrukov, A.~Magnin, M.~Shaposhnikov, and S.~Sibiryakov, ``{Higgs inflation:
  consistency and generalisations},''
  \href{http://dx.doi.org/10.1007/JHEP01(2011)016}{{\em JHEP} {\bfseries 01}
  (2011) 016}, \href{http://arxiv.org/abs/1008.5157}{{\ttfamily arXiv:1008.5157
  [hep-ph]}}.

\bibitem{Ema:2016dny}
Y.~Ema, R.~Jinno, K.~Mukaida, and K.~Nakayama, ``{Violent Preheating in
  Inflation with Nonminimal Coupling},''
  \href{http://dx.doi.org/10.1088/1475-7516/2017/02/045}{{\em JCAP} {\bfseries
  02} (2017) 045}, \href{http://arxiv.org/abs/1609.05209}{{\ttfamily
  arXiv:1609.05209 [hep-ph]}}.

\bibitem{Calmet:2013hia}
X.~Calmet and R.~Casadio, ``{Self-healing of unitarity in Higgs inflation},''
  \href{http://dx.doi.org/10.1016/j.physletb.2014.05.008}{{\em Phys. Lett. B}
  {\bfseries 734} (2014) 17--20},
  \href{http://arxiv.org/abs/1310.7410}{{\ttfamily arXiv:1310.7410 [hep-ph]}}.

\bibitem{Ema:2020zvg}
Y.~Ema, K.~Mukaida, and J.~van~de Vis, ``{Higgs Inflation as Nonlinear Sigma
  Model and Scalaron as its $\sigma$-meson},''
  \href{http://arxiv.org/abs/2002.11739}{{\ttfamily arXiv:2002.11739
  [hep-ph]}}.

\bibitem{Giudice:2010ka}
G.~F. Giudice and H.~M. Lee, ``{Unitarizing Higgs Inflation},''
  \href{http://dx.doi.org/10.1016/j.physletb.2010.10.035}{{\em Phys. Lett. B}
  {\bfseries 694} (2011) 294--300},
  \href{http://arxiv.org/abs/1010.1417}{{\ttfamily arXiv:1010.1417 [hep-ph]}}.

\bibitem{Ema:2017rqn}
Y.~Ema, ``{Higgs Scalaron Mixed Inflation},''
  \href{http://dx.doi.org/10.1016/j.physletb.2017.04.060}{{\em Phys. Lett. B}
  {\bfseries 770} (2017) 403--411},
  \href{http://arxiv.org/abs/1701.07665}{{\ttfamily arXiv:1701.07665
  [hep-ph]}}.

\bibitem{Barbon:2015fla}
J.~Barbon, J.~Casas, J.~Elias-Miro, and J.~Espinosa, ``{Higgs Inflation as a
  Mirage},'' \href{http://dx.doi.org/10.1007/JHEP09(2015)027}{{\em JHEP}
  {\bfseries 09} (2015) 027}, \href{http://arxiv.org/abs/1501.02231}{{\ttfamily
  arXiv:1501.02231 [hep-ph]}}.

\bibitem{Gorbunov:2018llf}
D.~Gorbunov and A.~Tokareva, ``{Scalaron the healer: removing the
  strong-coupling in the Higgs- and Higgs-dilaton inflations},''
  \href{http://dx.doi.org/10.1016/j.physletb.2018.11.015}{{\em Phys. Lett. B}
  {\bfseries 788} (2019) 37--41},
  \href{http://arxiv.org/abs/1807.02392}{{\ttfamily arXiv:1807.02392
  [hep-ph]}}.

\bibitem{He:2018gyf}
M.~He, A.~A. Starobinsky, and J.~Yokoyama, ``{Inflation in the mixed
  Higgs-$R^2$ model},''
  \href{http://dx.doi.org/10.1088/1475-7516/2018/05/064}{{\em JCAP} {\bfseries
  05} (2018) 064}, \href{http://arxiv.org/abs/1804.00409}{{\ttfamily
  arXiv:1804.00409 [astro-ph.CO]}}.

\bibitem{He:2018mgb}
M.~He, R.~Jinno, K.~Kamada, S.~C. Park, A.~A. Starobinsky, and J.~Yokoyama,
  ``{On the violent preheating in the mixed Higgs-$R^2$ inflationary model},''
  \href{http://dx.doi.org/10.1016/j.physletb.2019.02.008}{{\em Phys. Lett. B}
  {\bfseries 791} (2019) 36--42},
  \href{http://arxiv.org/abs/1812.10099}{{\ttfamily arXiv:1812.10099
  [hep-ph]}}.

\bibitem{Witten:1985cc}
E.~Witten, ``{Noncommutative Geometry and String Field Theory},''
  \href{http://dx.doi.org/10.1016/0550-3213(86)90155-0}{{\em Nucl. Phys. B}
  {\bfseries 268} (1986) 253--294}.

\bibitem{Witten:1986qs}
E.~Witten, ``{Interacting Field Theory of Open Superstrings},''
  \href{http://dx.doi.org/10.1016/0550-3213(86)90298-1}{{\em Nucl. Phys. B}
  {\bfseries 276} (1986) 291--324}.

\bibitem{Ohmori:2001am}
K.~Ohmori, ``{A Review on tachyon condensation in open string field
  theories},'' other thesis, 2, 2001.

\bibitem{Arefeva:2001ps}
I.~Arefeva, D.~Belov, A.~Giryavets, A.~Koshelev, and P.~Medvedev,
  \href{http://dx.doi.org/10.1142/9789812777317\_0001}{``{Noncommutative field
  theories and (super)string field theories},''} in {\em {11th Jorge Andre
  Swieca Summer School on Particle and Fields}}, pp.~1--163.
\newblock 11, 2001.
\newblock \href{http://arxiv.org/abs/hep-th/0111208}{{\ttfamily
  arXiv:hep-th/0111208}}.

\bibitem{Brekke:1988dg}
L.~Brekke, P.~G. Freund, M.~Olson, and E.~Witten, ``{Nonarchimedean String
  Dynamics},'' \href{http://dx.doi.org/10.1016/0550-3213(88)90207-6}{{\em Nucl.
  Phys. B} {\bfseries 302} (1988) 365--402}.

\bibitem{Vladimirov:1994wi}
V.~Vladimirov, I.~Volovich, and E.~Zelenov, {\em {p-adic analysis and
  mathematical physics}}, vol.~1.
\newblock 1994.

\bibitem{Dragovich:2017kge}
B.~Dragovich, A.~Y. Khrennikov, S.~Kozyrev, I.~Volovich, and E.~Zelenov,
  ``{$p$-Adic Mathematical Physics: The First 30 Years},''
  \href{http://dx.doi.org/10.1134/S2070046617020017}{{\em Anal. Appl.}
  {\bfseries 9} (2017) 87--121},
  \href{http://arxiv.org/abs/1705.04758}{{\ttfamily arXiv:1705.04758
  [math-ph]}}.

\bibitem{Kuzmin:1989sp}
Y.~Kuzmin, ``{THE CONVERGENT NONLOCAL GRAVITATION. (IN RUSSIAN)},'' {\em Sov.
  J. Nucl. Phys.} {\bfseries 50} (1989) 1011--1014.

\bibitem{Krasnikov:1987yj}
N.~Krasnikov, ``{NONLOCAL GAUGE THEORIES},''
  \href{http://dx.doi.org/10.1007/BF01017588}{{\em Theor. Math. Phys.}
  {\bfseries 73} (1987) 1184--1190}.

\bibitem{Tomboulis:1997gg}
E.~Tomboulis, ``{Superrenormalizable gauge and gravitational theories},''
  \href{http://arxiv.org/abs/hep-th/9702146}{{\ttfamily arXiv:hep-th/9702146}}.

\bibitem{Biswas:2011ar}
T.~Biswas, E.~Gerwick, T.~Koivisto, and A.~Mazumdar, ``{Towards singularity and
  ghost free theories of gravity},''
  \href{http://dx.doi.org/10.1103/PhysRevLett.108.031101}{{\em Phys. Rev.
  Lett.} {\bfseries 108} (2012) 031101},
  \href{http://arxiv.org/abs/1110.5249}{{\ttfamily arXiv:1110.5249 [gr-qc]}}.

\bibitem{Biswas:2016egy}
T.~Biswas, A.~S. Koshelev, and A.~Mazumdar, ``{Consistent higher derivative
  gravitational theories with stable de Sitter and anti--de Sitter
  backgrounds},'' \href{http://dx.doi.org/10.1103/PhysRevD.95.043533}{{\em
  Phys. Rev. D} {\bfseries 95} no.~4, (2017) 043533},
  \href{http://arxiv.org/abs/1606.01250}{{\ttfamily arXiv:1606.01250 [gr-qc]}}.

\bibitem{Koshelev:2017ebj}
A.~S. Koshelev, K.~Sravan~Kumar, L.~Modesto, and L.~a. Rachwa\l, ``{Finite
  quantum gravity in dS and AdS spacetimes},''
  \href{http://dx.doi.org/10.1103/PhysRevD.98.046007}{{\em Phys. Rev. D}
  {\bfseries 98} no.~4, (2018) 046007},
  \href{http://arxiv.org/abs/1710.07759}{{\ttfamily arXiv:1710.07759
  [hep-th]}}.

\bibitem{Ostro:1850}
M.~Ostrogradsky, ``Memoires sur les equations differentielles relatives au
  probleme des isoperimetres,'' {\em Mem. Ac. St. Petersbourg} {\bfseries VI}
  no.~4, (1850) 385.

\bibitem{Coleman:1968wh}
S.~Coleman, ``{Resonance poles and resonance multipoles},''.

\bibitem{Hawking:2000bb}
S.~Hawking, T.~Hertog, and H.~Reall, ``{Trace anomaly driven inflation},''
  \href{http://dx.doi.org/10.1103/PhysRevD.63.083504}{{\em Phys. Rev. D}
  {\bfseries 63} (2001) 083504},
  \href{http://arxiv.org/abs/hep-th/0010232}{{\ttfamily arXiv:hep-th/0010232}}.

\bibitem{Anselmi:2018kgz}
D.~Anselmi, ``{Fakeons And Lee-Wick Models},''
  \href{http://dx.doi.org/10.1007/JHEP02(2018)141}{{\em JHEP} {\bfseries 02}
  (2018) 141}, \href{http://arxiv.org/abs/1801.00915}{{\ttfamily
  arXiv:1801.00915 [hep-th]}}.

\bibitem{Starobinsky:1980te}
A.~A. Starobinsky, ``{A New Type of Isotropic Cosmological Models Without
  Singularity},'' \href{http://dx.doi.org/10.1016/0370-2693(80)90670-X}{{\em
  Adv. Ser. Astrophys. Cosmol.} {\bfseries 3} (1987) 130--133}.

\bibitem{Starobinsky:1981vz}
A.~A. Starobinsky, ``{NONSINGULAR MODEL OF THE UNIVERSE WITH THE QUANTUM
  GRAVITATIONAL DE SITTER STAGE AND ITS OBSERVATIONAL CONSEQUENCES},'' in {\em
  {Second Seminar on Quantum Gravity}}, pp.~103--128.
\newblock 1, 1981.

\bibitem{Starobinsky:1983zz}
A.~A. Starobinsky, ``{The Perturbation Spectrum Evolving from a Nonsingular
  Initially De-Sitter Cosmology and the Microwave Background Anisotropy},''
  {\em Sov. Astron. Lett.} {\bfseries 9} (1983) 302.

\bibitem{Koshelev:2007fi}
A.~S. Koshelev, ``{Non-local SFT Tachyon and Cosmology},''
  \href{http://dx.doi.org/10.1088/1126-6708/2007/04/029}{{\em JHEP} {\bfseries
  04} (2007) 029}, \href{http://arxiv.org/abs/hep-th/0701103}{{\ttfamily
  arXiv:hep-th/0701103}}.

\bibitem{Koshelev:2010bf}
A.~S. Koshelev and S.~Y. Vernov, ``{Analysis of scalar perturbations in
  cosmological models with a non-local scalar field},''
  \href{http://dx.doi.org/10.1088/0264-9381/28/8/085019}{{\em Class. Quant.
  Grav.} {\bfseries 28} (2011) 085019},
  \href{http://arxiv.org/abs/1009.0746}{{\ttfamily arXiv:1009.0746 [hep-th]}}.

\bibitem{Arefeva:2008zru}
I.~Aref'eva and A.~Koshelev, ``{Cosmological Signature of Tachyon
  Condensation},'' \href{http://dx.doi.org/10.1088/1126-6708/2008/09/068}{{\em
  JHEP} {\bfseries 09} (2008) 068},
  \href{http://arxiv.org/abs/0804.3570}{{\ttfamily arXiv:0804.3570 [hep-th]}}.

\bibitem{1106.3439}
{Lasse Rempe-Gillen}, ``{Hyperbolic entire functions with full hyperbolic
  dimension and approximation by Eremenko-Lyubich functions},''
  \href{http://dx.doi.org/{10.1112/plms/pdt048}}{{\em Proc. London Math. Soc.}
  {\bfseries 108} no.~5, ({2011}) 1193--1225},
  \href{http://arxiv.org/abs/{1106.3439}}{{\ttfamily arXiv:{1106.3439}
  [math]}}.

\bibitem{Maldacena:2002vr}
J.~M. Maldacena, ``{Non-Gaussian features of primordial fluctuations in single
  field inflationary models},''
  \href{http://dx.doi.org/10.1088/1126-6708/2003/05/013}{{\em JHEP} {\bfseries
  05} (2003) 013}, \href{http://arxiv.org/abs/astro-ph/0210603}{{\ttfamily
  arXiv:astro-ph/0210603}}.

\bibitem{Creminelli:2004yq}
P.~Creminelli and M.~Zaldarriaga, ``{Single field consistency relation for the
  3-point function},''
  \href{http://dx.doi.org/10.1088/1475-7516/2004/10/006}{{\em JCAP} {\bfseries
  10} (2004) 006}, \href{http://arxiv.org/abs/astro-ph/0407059}{{\ttfamily
  arXiv:astro-ph/0407059}}.

\bibitem{Chen:2006nt}
X.~Chen, M.-x. Huang, S.~Kachru, and G.~Shiu, ``{Observational signatures and
  non-Gaussianities of general single field inflation},''
  \href{http://dx.doi.org/10.1088/1475-7516/2007/01/002}{{\em JCAP} {\bfseries
  01} (2007) 002}, \href{http://arxiv.org/abs/hep-th/0605045}{{\ttfamily
  arXiv:hep-th/0605045}}.

\bibitem{Koshelev:2020foq}
A.~S. Koshelev, K.~Sravan~Kumar, A.~Mazumdar, and A.~A. Starobinsky,
  ``{Non-Gaussianities and tensor-to-scalar ratio in non-local $R^2$-like
  inflation},'' \href{http://arxiv.org/abs/2003.00629}{{\ttfamily
  arXiv:2003.00629 [hep-th]}}.

\bibitem{Pius:2016jsl}
R.~Pius and A.~Sen, ``{Cutkosky rules for superstring field theory},''
  \href{http://dx.doi.org/10.1007/JHEP10(2016)024}{{\em JHEP} {\bfseries 10}
  (2016) 024}, \href{http://arxiv.org/abs/1604.01783}{{\ttfamily
  arXiv:1604.01783 [hep-th]}}. [Erratum: JHEP 09, 122 (2018)].

\bibitem{Bezrukov:2014ipa}
F.~Bezrukov, J.~Rubio, and M.~Shaposhnikov, ``{Living beyond the edge: Higgs
  inflation and vacuum metastability},''
  \href{http://dx.doi.org/10.1103/PhysRevD.92.083512}{{\em Phys. Rev. D}
  {\bfseries 92} no.~8, (2015) 083512},
  \href{http://arxiv.org/abs/1412.3811}{{\ttfamily arXiv:1412.3811 [hep-ph]}}.

\bibitem{Langacker:2010zza}
P.~Langacker, {\em {The standard model and beyond}}.
\newblock 2, 2010.

\bibitem{Faddeev:1980be}
L.~Faddeev and A.~Slavnov, {\em {GAUGE FIELDS. INTRODUCTION TO QUANTUM
  THEORY}}, vol.~50.
\newblock 1980.

\bibitem{Bardeen:1995kv}
W.~A. Bardeen, ``{On naturalness in the standard model},'' in {\em {Ontake
  Summer Institute on Particle Physics}}.
\newblock 8, 1995.

\bibitem{Shaposhnikov:2008xi}
M.~Shaposhnikov and D.~Zenhausern, ``{Quantum scale invariance, cosmological
  constant and hierarchy problem},''
  \href{http://dx.doi.org/10.1016/j.physletb.2008.11.041}{{\em Phys. Lett. B}
  {\bfseries 671} (2009) 162--166},
  \href{http://arxiv.org/abs/0809.3406}{{\ttfamily arXiv:0809.3406 [hep-th]}}.

\bibitem{Shaposhnikov:2018jag}
M.~Shaposhnikov and A.~Shkerin, ``{Gravity, Scale Invariance and the Hierarchy
  Problem},'' \href{http://dx.doi.org/10.1007/JHEP10(2018)024}{{\em JHEP}
  {\bfseries 10} (2018) 024}, \href{http://arxiv.org/abs/1804.06376}{{\ttfamily
  arXiv:1804.06376 [hep-th]}}.

\bibitem{Koshelev:2020xby}
A.~S. Koshelev, K.~S. Kumar, and A.~A. Starobinsky, ``{Analytic infinite
  derivative gravity, $R^2$-like inflation, quantum gravity and CMB},''
  \href{http://arxiv.org/abs/2005.09550}{{\ttfamily arXiv:2005.09550
  [hep-th]}}.

\end{thebibliography}

\providecommand{\href}[2]{#2}\begingroup\raggedright\endgroup

\end{document}